\documentclass{article}
\usepackage{imakeidx}
\usepackage[toc,page]{appendix}
\pdfoutput=1
\usepackage{graphicx}
\usepackage{epstopdf}
\usepackage{amsmath, amsthm, latexsym}
\usepackage{amssymb}
\usepackage{slashed}

\usepackage[nottoc]{tocbibind}

\usepackage{color}

\usepackage[small, bf]{caption}
\usepackage{subfigure}
\usepackage{hyperref}

\newtheorem{proposition}{Proposition}
\newtheorem{remark}{Remark}
\newtheorem{exercise}{Exercise}

\usepackage{xcolor}
\hypersetup{colorlinks, linkcolor={red!70!black}, citecolor={cyan}, urlcolor={blue!80!black}}

\numberwithin{equation}{section}

\makeindex[intoc]

\begin{document}

\begin{titlepage}\begin{center}
\textbf{\Large{The Dirac equation in general relativity}}\\\vspace{6mm}
\textbf{\large {A guide for calculations}}\\\vspace{6mm}
\textbf{\normalsize Peter Collas}\footnote{Department of Physics and Astronomy, California State University, Northridge, Northridge, CA 91330-8268. Email: peter.collas@csun.edu.}
\textbf{\normalsize and David Klein}\footnote{Department of Mathematics and Interdisciplinary Research Institute for the Sciences, California State University, Northridge, Northridge, CA 91330-8313. Email: david.klein@csun.edu.}\\
\end{center}

\begin{center} (September 7, 2018) \end{center}
\vspace{2cm}

\begin{abstract}

\noindent In these informal lecture notes we outline different approaches used in doing calculations involving the Dirac equation in curved spacetime.  We have tried to clarify the subject by carefully pointing out the various conventions used and by including several examples from textbooks and the existing literature.  In addition some basic material has been included in the appendices.  It is our hope that graduate students and other researchers will find these notes useful.

\end{abstract}

\end{titlepage}
\hypersetup{linktocpage}
\tableofcontents

\newpage

\section[The spinorial covariant derivative]{The spinorial covariant derivative} \label{SCD}

\subsection[The Fock-Ivanenko coefficients]{The Fock-Ivanenko coefficients}\label{FI I}

\noindent In the calculations that follow we will specify our metric signature\index{metric!signature} as needed.  We adopt the convention that upper case latin indices run over $(0, 1, 2, 3)$.  Also we shall adopt Planck units so that, $c=G=\hbar=1$.  A review of the gamma matrix\index{gamma matrices!in Minkowski spacetime} representations and the Dirac equation\index{Dirac equation!in Minkowski spacetime} in Minkowski spacetime is given in Appendices \ref{Gam} and \ref{Dir}.\\

\noindent Fermion fields are described by spinors $\psi(x)$,\index{spinor!field} and in order to accommodate spinors in general relativity we need the tetrad formalism.  The tetrad formalism is briefly reviewed in Appendix \ref{Tet}, where apart from the standard material we have included some additional material relevant to spinors \cite{L76}.\\

\noindent From the basic tetrad expression, Eq. \eqref{T2} in Appendix \ref{Tet}, namely,
\begin{equation}
\label{1.0}
\eta_{AB}=e_A{}^{\alpha}\,e_B{}^{\beta}\,g_{\alpha\beta}\,,
\end{equation}

\noindent we see that $\eta_{00}=e_0{}^{\alpha}\,e_0{}^{\beta}\,g_{\alpha\beta}$, and thus by definition the tetrad vector\index{tetrad!vectors} $e_0{}$ is a velocity field at least momentarily tangent to a timelike path.  This is what Schutz \cite{S09} refers to as the ``momentarily comoving reference frame'' (MCRF) and it is in this sense that our choice of a tetrad vector set,  $e_A{}^\alpha{}$, determines the frame we shall refer to as the  \textit{the reference frame for the Dirac particle}\index{reference frame!of Dirac particle}, or \textit{the particle frame}\index{particle frame} for short.\\

\noindent For example in reference \cite{P80}, Sec. VI, Parker considers a freely falling hydrogenic atom, with its nucleus on the geodesic (an approximation), and constructs approximate Fermi coordinates along the chosen geodesic.  The corresponding tetrad is referred to as \textit{the proper frame}\index{frame!proper}.  In general the choice of a tetrad set may be dictated for reasons of convenience and one should read the comments in Remark \ref{pfid} and keep in mind the fact that from a given tetrad one can obtain an infinity of tetrads related to each other by local Lorentz transformations (see Sec. \ref{1DM}).\\

\noindent In general relativity the spinors, $\psi(x)$, are sections of the spinor bundle\index{spinor!bundle}.  We limit ourselves to presenting the bare essentials required for calculations, and on clarifying the different sign conventions related to the definition of the spinorial covariant derivative,\index{spinorial covariant derivative} the spinor affine connection,\index{spinor!affine connection} $\Gamma_\mu{}$, and Fock-Ivanenko coefficients\index{Fock-Ivanenko coefficients} $\Gamma_C{}$.\\

\noindent Each component of a spinor transforms as a scalar function under general coordinate transformations,\index{coordinate trasformations} so this kind of transformation is straightforward.  However, the transformation of spinors under tetrad rotations requires additional formalism.\\

\noindent If we change from an initial set of tetrad vector fields, $h_A{}$, to another set, $e_A{}$, then the new tetrad vectors can be expressed as linear combinations of the old as shown below
\begin{equation}
\label{FI1}
e_A{}^\mu{}=\Lambda_A{}^B{}h_B{}^\mu{}\,,
\end{equation}

\noindent We show in Appendix \ref{V&LB} that $\Lambda$ is a Lorentz matrix.\index{Lorentz!matrix}  So in the context of general relativity \textit{the Lorentz group\index{Lorentz!group} is the group of tetrad\index{tetrad} rotations} \cite{L76}.  We also remark that the $\Lambda$ matrices are in general spacetime-dependent and we refer to them as local Lorentz transformations\index{Lorentz!transformation}.\\

\noindent In order to write the Dirac equation\index{Dirac equation!in curved spacetime} in general relativity, we also need to introduce the spacetime dependent matrices\index{gamma matrices!spacetime dependent} $\bar{\gamma}^\alpha{}(x)$.  The $\bar{\gamma}^\alpha{}$ matrices are related to the constant special relativity gamma matrices\index{gamma matrices!in Minkowski spacetime}, $\gamma^A{}$, by the relation
\begin{equation}
\label{FI1b}
\bar{\gamma}^\alpha{}(x):=e_A{}^\alpha{}(x)\gamma^A{}\,,
\end{equation}

\noindent Using Eq. \eqref{T2} we can now relate the anti-commutators below,
\begin{align}
&\{\gamma^{A},\gamma^{B}\}=\varepsilon\, 2\eta^{AB}I\,,\label{FI1c}\\\nonumber\\
&\{\bar{\gamma}^\alpha{}(x),\bar{\gamma}^\beta{}(x)\}=\varepsilon\,2g^{\alpha\beta}I\,,\label{FI1a}
\end{align}

\noindent where $\varepsilon=\pm 1$.  We note that the matrices in Eq. \eqref{FI1a} anticommute for $\alpha\neq\beta$, only if the metric is diagonal.\\

\noindent A spinor, $\psi$, may be defined as a quantity that transforms as
\begin{equation}
\label{1a}
\tilde{\psi}_{e}=L\,\psi_{h}\,,
\end{equation}

\noindent where $L=L(x)$ is the spacetime-dependent spinor representative of a tetrad rotation $\Lambda=\Lambda(x)$ \cite{L76}.  We initially follow the sign conventions of references \cite{CL76} - \cite{F89}, and although all these references use the metric signature $(+,-,-,-)$, we shall maintain, wherever possible, greater generality.\\

\noindent The derivative of a spinor does not transform like a spinor since
\begin{equation}
\label{1b}
\tilde{\psi},_{\mu}=L\,\psi,_{\mu}+L,_{\mu}\psi\,.
\end{equation}

\noindent Therefore we define the covariant derivative of a spinor by the expression,
\begin{equation}
\label{1c}
D_\mu{}\,\psi=I\psi,_{\mu}+\Gamma_\mu{}\,\psi\,,
\end{equation}

\noindent with the spinor affine connection\index{connection}, $\Gamma_\mu{}$ to be determined.  The connection $\Gamma_{\mu}$ is a matrix, actually four matrices, that is, $(\Gamma_{\mu})_a{}^b{}$.  We require that,
\begin{equation}
\tilde{D}_\mu{}\,\tilde{\psi}=LD_\mu{}\,\psi\,,\label{1c1}
\end{equation}

\noindent where
\begin{equation}
\tilde{D}_\mu{}\,\tilde{\psi}=I\tilde{\psi},_{\mu}+\tilde{\Gamma}_\mu{}\,\tilde{\psi}\,.\label{1c2}
\end{equation}

\noindent Eq. \eqref{1c1} is satisfied if we let
\begin{equation}
\label{1d}
\tilde{\Gamma}_\mu{}=L\,\Gamma_\mu{}L^{-1}-L,_{\mu}L^{-1},
\end{equation}

\noindent since then
\begin{align}
\tilde{D}_\mu{}\tilde{\psi}&=\partial_\mu{}(L\psi)+\tilde{\Gamma}_\mu{}\,\tilde{\psi}\,,\label{1e}\\
&=(L,_{\mu})\psi+L\psi,_{\mu}+L\Gamma_\mu{}\psi-(L,_{\mu})\psi\,.\label{1f}\\
&=L(I\psi,_{\mu}+\,\Gamma_\mu{}\,\psi)\,.
\end{align}

\noindent With a slight abuse of notation we write the spinor covariant derivative \index{spinor covariant derivative}acting on a spinor $\psi(x)$ as
\begin{equation}
\label{1}
D_\mu{}\,\psi=\left(I\partial_\mu{}+\Gamma_\mu{}\right)\psi:=\left(\partial_\mu{}+\Gamma_\mu{}\right)\psi\,,
\end{equation}

\noindent  where we may omit the identity matrix factor $I$ in the second part of Eq. \eqref{1}.  We now proceed to deduce the expression for $\Gamma_\mu{}$.\\

\noindent Under the assumption that the operator $D_\mu{}$ is a connection and therefore a derivation \index{derivation}(i.e., satisfies the product rule for tensor products), it may be extended as an operator on a matrix-valued field $M$, \cite{F89}, \cite{B96}.  By writing $M$ as a linear combination of tensor products of vectors with co-vectors, a calculation shows that
\begin{equation}
\label{1g}
D_{\mu}M=\nabla_{\mu}M+\left[\Gamma_{\mu},M\right].
\end{equation}

\noindent  In particular, if $M=I$, then $D_\mu{}M=0$.  We now impose the additional requirement that the derivative, $D_\mu{}$, is \textit{metric compatible}, i.e.,
\begin{equation}
\label{2}
D_\mu{}\,g^{\alpha\beta}I=0\,,
\end{equation}

\noindent where $g^{\alpha\beta}$ in this expression is understood to be a scalar (the element of a matrix) rather than a matrix.  Recalling Eq. \eqref{FI1a},
\begin{equation}
\label{8}
\varepsilon\,2g^{\alpha\beta}I=\{\bar{\gamma}^\alpha{}(x),\bar{\gamma}^\beta{}(x)\}\,,
\end{equation}

\noindent we see that Eq. \eqref{2} is equivalent to
\begin{equation}
\label{9}
D_\mu{}\left(\{\bar{\gamma}^\alpha{}(x),\bar{\gamma}^\beta{}(x)\}\right)=0\,,
\end{equation}

\noindent and a \textit{sufficient} condition for the above equation is
\begin{equation}
\label{10}
D_\mu{}\bar{\gamma}^\nu{(x)}=0\,.
\end{equation}

\noindent The operator $D_\mu$ of Eq. \eqref{1g} acting on $\bar{\gamma}^\nu$ is,   
\begin{equation}\label{11}
D_\mu{}\bar{\gamma}^\nu{}=\nabla_\mu{}\bar{\gamma}^\nu{}+\left[\Gamma_\mu{},\bar{\gamma}^\nu{}\right].
\end{equation}

\noindent Thus Eq. \eqref{10} is
\begin{equation}
\label{12}
D_\mu{}\bar{\gamma}^\nu{}=\bar{\gamma}^\nu{}_{,\mu}+\Gamma^\nu{}_{\lambda\mu}\bar{\gamma}^\lambda{}+\Gamma_\mu{}\bar{\gamma}^\nu{}-\bar{\gamma}^\nu{}\Gamma_\mu{}=0\,.
\end{equation}

\noindent We refer the reader to Appendix \ref{SigCartan}, for further details of the effect of signature choices on the $\Gamma_\mu{}$ and the Dirac equation.

\begin{remark}\label{sig2}\index{Remarks} Reference \textup{\cite{L76}, Eq. (13.27)}, for example, has a $(-)$ sign in front of the commutator in \textup{Eq}. \eqref{11} thus, effectively, changing the sign of $\Gamma_\mu{}$. This is compensated for by writing $\left(\partial_\mu{}-\Gamma_\mu{}\right)$ in Eq. \eqref{1}.\end{remark}

\noindent We now introduce the \textit{spin} connection\index{spin connection} (coefficients) $\omega^A{}_{B\mu}$ by the relation below (e.g., see \cite{PT09}, pp.222-224, and \cite{C04}, p. 487),
\begin{equation}
\label{FI1a}
\omega^A{}_{B\mu}:=-e_B{}^\nu{}\left(\partial_{\mu}e^A{}_\nu{}-\Gamma^\lambda{}_{\mu\nu}e^A{}_\lambda{}\right)\,.
\end{equation}

\begin{equation}
\label{FI1}
\omega_{AB\mu}=e_{A\beta}\,\nabla_{\mu}\,e_B{}^{\beta}=g_{\beta\alpha}e_A{}^\alpha{}\,\nabla_{\mu}\,e_B{}^{\beta}=\eta_{AC}\,e^C{}_\beta{}\,\nabla_{\mu}\,e_B{}^{\beta}\,.
\end{equation}

\noindent  We see from the second (or third) equality in Eq. \eqref{FI1} that the metric signature \index{metric!signature} will affect the signs of the $\omega_{AB\mu}$.\\

\begin{exercise}\label{SpC}\index{Exercises} Obtain \textup{Eq}. \eqref{FI1a} using \textup{Eq} \eqref{FI1} and the result of Proposition \textup{\ref{antisym}} below. \textup{(}Hint:  Pay attention to the ordering of the indices $A$ and $B$ in Eqs. \eqref{FI1a} and \eqref{FI1}.\textup{)}
\end{exercise}

\begin{proposition}\label{antisym}\index{Propositions}The $\omega_{AB\mu}$ are antisymmetric in $A$ and $B$. \end{proposition}

\noindent \textit{Proof}: Recall Eq. \eqref{T2}, then we have that (cf. \cite{C04}, p.489),
\begin{align}
\nabla_{\mu}\,\eta_{AB}=e_B{}^{\beta}\left(\nabla_{\mu}\,e_A{}^{\alpha}\right)g_{\alpha\beta}+e_A{}^{\alpha}\left(\nabla_{\mu}\,e_B{}^{\beta}\right)g_{\alpha\beta}&=0\,,\nonumber\\
e_{B\alpha}\left(\nabla_{\mu}\,e_A{}^{\alpha}\right)+e_{A\beta}\left(\nabla_{\mu}\,e_B{}^{\beta}\right)&=0\,,\nonumber\\
\omega_{BA\mu}+\omega_{AB\mu}&=0\,.\label{FI9}
\end{align}

\begin{remark}\label{term}\index{Remarks}The $\omega_{AB\mu}$ are also written as $\omega_{\mu AB}$, \textup{\cite{PT09}}.  Lord refers to $\nabla_{\mu}\,e_B{}^{\beta}$ as the Ricci rotation coefficients while in our nomenclature the Ricci rotation coefficients are given by Eq. \eqref{FI21}, namely their components along the tetrad field $e_C{}^{\mu}$.  Our point here is that the terminology varies a little in the literature and care is required. \end{remark}

\noindent Using Eq. \eqref{FI1b} and \eqref{FI1c} in Eq. \eqref{11}, one can show that the $\Gamma_\mu{}$ below satisfies Eq. \eqref{12} and hence \eqref{2},
\begin{equation}
\label{FI10}
\Gamma_{\mu}=\frac{\varepsilon}{4}\,\omega_{AB\mu}\,\gamma^{A}\gamma^{B}=\frac{\varepsilon}{2}\,\omega_{AB\mu}\,\Sigma^{AB}\,,
\end{equation}

\noindent where
\begin{equation}
\label{FI11}
\Sigma^{AB}=\frac{1}{4}\left[\gamma^A{},\gamma^B{}\right]\,.
\end{equation}

\noindent We have included an $\varepsilon$ factor in the expression for $\Gamma_\mu{}$ in order take into account the choice made in Eq. \eqref{FI1c}.  The factor of $1/4$ in $\Gamma_\mu{}$, Eq. \eqref{FI10}, compensates for the factor of $2$ in Eq. \eqref{FI1c} and thus is dimension independent.  The reader may consult Fursaev and Vassilevich \cite{FV11}, p.16 for a different line of reasoning.

\begin{remark}\label{ParkerToms}\index{Remarks} We caution the reader that Parker and Toms \textup{\cite{PT09}} define a $\Gamma_\mu{}$ on p. 145 which is our Eq. \eqref{FI10} with $\varepsilon=-1$ and a $B_\mu{}$ on p. 228, which is our Eq. \eqref{FI10} with $\varepsilon=+1$. No problem arises since the corresponding Dirac equations also differ in the appropriate sign.\end{remark}

\noindent The \textit{Fock-Ivanenko} coefficients,\index{Fock-Ivanenko coefficients} $\Gamma_{C}$, are given by
\begin{equation}
\label{FI12b}
\Gamma_{C}=e_C{}^{\mu}\,\Gamma_{\mu}\,,
\end{equation}

\noindent thus we may write
\begin{equation}
\label{FI12c}
D_{C}\,\psi=\left(e_{C}+\Gamma_{C}\right)\psi\,.
\end{equation}

\noindent Finally, for a free spin $1/2$ particle of mass $m$ we have the Dirac equation\index{Dirac equation!in curved spacetime} in curved spacetime,
\begin{equation}
\label{FI13}
i\gamma^{C}D_C{}\,\psi-m\psi=0\,.
\end{equation}

\noindent Using Eqs. \eqref{FI1b}, \eqref{1}, and \eqref{FI12b}, we may also write Eq. \eqref{FI13} in the form
\begin{align}
i\bar{\gamma}^{\mu}(\partial_\mu{}+\Gamma_\mu{})\,\psi-m\psi&=0\,,\label{FI13a}\\\nonumber\\
i\bar{\gamma}^{\mu}D_\mu{}\psi-m \psi&=0\,.\label{FI13b}
\end{align}

\begin{remark}\label{operator}\index{Remarks} We note that the $e_C{}=e_C{}^\gamma{}\,\partial_\gamma{}$, in \textup{Eqs.} \eqref{FI12c}, \eqref{FI13}, is regarded as a differential operator, thus for our four component spinor $\psi$, we have, after re-inserting the identity matrix $I$,\end{remark}
\begin{equation}
\label{FI14c}
e_C{}I\,\psi=\left(\begin{array}{c}e_C{}^\gamma{}\,\partial_\gamma{}\,\psi_1{}\\\\
e_C{}^\gamma{}\,\partial_\gamma{}\,\psi_2{}\\\\
e_C{}^\gamma{}\,\partial_\gamma{}\,\psi_3{}\\\\
e_C{}^\gamma{}\,\partial_\gamma{}\,\psi_4{}
\end{array}\right).
\end{equation}\\

\begin{proposition}\label{parallel}\index{Propositions}If the tetrad $e_B{}^\beta{}$ in Eq. \eqref{FI1} is parallel\index{parallel} along a path with tangent $e_0{}^\mu{}$, then it is a Fermi tetrad\index{tetrad!Fermi}, \textup{(}see Eq. \eqref{FT5}\textup{)} and it turns out that the Fock-Ivanenko coefficient\index{Fock-Ivanenko coefficients} $\Gamma_{0}=0$. \end{proposition}

\noindent \textit{Proof}:  Using Eqs. \eqref{FI1}, \eqref{FI10}, we obtain the expression below for Eq. \eqref{FI12b}, 
\begin{equation}
\label{FI14d}
\Gamma_{C}=\frac{\varepsilon}{4}e_{A\beta}\left(e_C{}^\mu{}\nabla_{\mu}e_B{}^\beta{}\right)\gamma^{A}\gamma^{B}\,,
\end{equation}

\noindent thus if $e_B{}^\beta{}$ is parallel,\index{parallel} the terms in parentheses vanish for $C=0$.

\begin{remark}\label{Gamma}\index{Remarks} Conversely, if $\Gamma_{0}\neq0$, then $e_B{}^\beta{}$ is not parallel.\index{parallel}  However, it may happen that $\Gamma_{0}=0$, while $e_B{}^\beta{}$ is not parallel.\index{parallel} \end{remark}

\subsection[The Ricci rotation coefficient approach]{The Ricci rotation coefficient approach}\label{Soleng}

\noindent We now give an alternative, and possibly more efficient, way of calculating the Fock-Ivanenko coefficients.\index{Fock-Ivanenko coefficients}  We remark again that the terminology and definitions of some the quantities below vary in the literature and one has to be very careful.  We work in the tetrad frame \index{tetrad!frame}and define the \textit{structure coefficients} \index{structure coefficients} (or \textit{structure constants}\index{structure constants}), $C^D{}_{AB}$, by the relations below, \cite{R09}, \cite{LL03}, \cite{C92}.
\begin{equation}
\label{FI14e}
de^D{}=-\frac{1}{2}\,C^D{}_{AB}\,e^{A}\wedge e^{B}\,.
\end{equation}

\noindent An equivalent expression is
\begin{equation}
\label{FI15}
\left[e_A{},e_B{}\right]^{\gamma}=e_A{}^\alpha{}\,\partial_\alpha{}\,e_B{}^\gamma{}-e_B{}^{\beta}\,\partial_\beta{}\,e_A{}^\gamma{}=C^D{}_{AB}\,e_D{}^{\gamma}\,,
\end{equation}

\noindent while a most convenient expression is
\begin{equation}
\label{FI16}
C^D{}_{AB}=\left(e^D{}_{\alpha,\,\beta}-e^D{}_{\beta,\,\alpha}\right)e_A{}^\alpha{}e_B{}^\beta{}\,.
\end{equation}

\noindent We derive below Eq. \eqref{FI16} from Eq. \eqref{FI15}.  We begin by multiplying both sides of Eq. \eqref{FI15} by $e^C{}_\gamma{}$ and using Eq. \eqref{T1b}. Thus
\begin{equation}
\label{FI17}
e_A{}^\alpha{}\,\left(\partial_\alpha{}\,e_B{}^\gamma{}\right)e^C{}_\gamma{}-e_B{}^{\beta}\,\left(\partial_\beta{}\,e_A{}^\gamma{}\right)e^C{}_\gamma{}=C^D{}_{AB}\,\delta_D{}^C{}=C^C{}_{AB}\,.
\end{equation}

\noindent We also have the identities
\begin{align}
\partial_\alpha{}\left(e_B{}^\gamma{}e^C{}_\gamma{}\right)&=\partial_\alpha{}\delta_B{}^C{}=\left(\partial_\alpha{}e_B{}^\gamma{}\right)e^C{}_\gamma{}+e_B{}^\gamma{}\partial_\alpha{}e^C{}_\gamma=0\,,\label{FI17a}\\
\partial_\beta{}\left(e_A{}^\gamma{}e^C{}_\gamma{}\right)&=\partial_\beta{}\delta_A{}^C{}=\left(\partial_\beta{}e_A{}^\gamma{}\right)e^C{}_\gamma{}+e_A{}^\gamma{}\partial_\beta{}e^C{}_\gamma=0\,.\label{FI17b}
\end{align}

\noindent Therefore Eq. \eqref{FI17} is
\begin{equation}
\label{FI17c}
-e_A{}^\alpha{}e_B{}^\gamma{}e^C{}_{\gamma,\,\alpha}+e_B{}^\beta{}e_A{}^\gamma{}e^C{}_{\gamma,\,\beta}=C^C{}_{AB}\,,
\end{equation}

\noindent which, after relabeling the dummy indices, reduces to Eq. \eqref{FI16}.\\

\noindent Clearly
\begin{equation}
\label{FI18}
C^D{}_{AB}=-\,C^D{}_{BA}\,,
\end{equation}

\noindent and we have,
\begin{align}
C_{ABC}&=\eta_{AD}C^D{}_{BC}\,,\label{FI19}\\
C_{ABC}&=-C_{ACB}\,.\label{FI20}
\end{align}

\noindent Finally we give two expressions for the \textit{Ricci rotation coefficients},\index{Ricci rotation coefficients} $\Gamma_{ABC}$.  The $\Gamma_{ABC}$ are related to the $\omega_{AB\mu}$, defined in Eq. \eqref{FI1}, by the relation Eq. \eqref{FI21} below, where the metric signature affects the $\omega_{AB\mu}$ and hence the $\Gamma_{ABC}$.
\begin{equation}
\label{FI21}
\Gamma_{ABC}=\omega_{AB\mu}\,e_C{}^\mu{}\,.
\end{equation}

\noindent Another expression for the $\Gamma_{ABC}$, which is written here \textit{specifically for the metric signature} $(+,-,-,-)$ is 
\begin{equation}
\label{FI22}
\Gamma_{ABC}=-\frac{1}{2}\left(C_{ABC}+C_{BCA}-C_{CAB}\right)\,.
\end{equation}

\noindent Eqs. \eqref{FI21} and \eqref{FI22} agree with the definitions in ref. \cite{C92}, Eqs. (253), p. 37, and (272), p. 39.  For metric signature \index{metric!signature}$(-,+,+,+)$ one has to change the overall sign in Eq. \eqref{FI22}, \textit{in addition note} that the $C$'s in Eq. \eqref{FI15} are the negatives of the $C$'s defined in \cite{Cv1.8}, Eq. (2.11) (see  also Remark \ref{Cartan FI1}).  From Eq. \eqref{FI21} we see that
\begin{equation}
\label{FI23}
\Gamma_{ABC}=-\,\Gamma_{BAC}\,.
\end{equation}

\noindent It is important to keep in mind the difference in index antisymmetry in Eqs. \eqref{FI20} and \eqref{FI23}.  Finally, using Eqs. \eqref{FI10}, \eqref{FI12b}, and\eqref{FI21}, we may now express the Fock-Ivanenko coefficients \index{Fock-Ivanenko coefficients}in terms of the $\Gamma_{ABC}$,
\begin{equation}
\label{FI24}
\Gamma_C{}=\frac{\varepsilon}{4}\,\Gamma_{ABC}\,\gamma^A{}\gamma^B{}\,.
\end{equation}

\noindent We have used as references, \cite{R09}, \cite{LL03}, and \cite{Cv1.8}. (Ref. \cite{R09} has some misprints in Sec. 11.4.)  If one were to adopt the definitions and terminology of Soleng \cite{Cv1.8}, one would have the advantage of being able to use the Mathematica package \textsc{Cartan}\index{Cartan@\textsc{Cartan}} to calculate all of these quantities (symbolically) by computer.

\begin{remark}\label{rep}\index{Remarks} Apart from being tetrad-dependent, it is clear from the above derivation, that the sign of the Ricci rotation coefficients,\index{Ricci rotation coefficients} $\Gamma_{ABC}$, will depend on the metric signature\index{metric!signature} since $C_{ABC}=\eta_{AD}C^D{}_{BC}$ \textup{(}we add that certain authors, e.g., \textup{\cite{LL03}}, define their spin connection Eq. \eqref{FI1} with the opposite sign\textup{)}.  Furthermore, it is after Eq. \eqref{FI24}, that is, when we write Eq. \eqref{FI13},that we have to choose a metric compatible representation of the $\gamma$ matrices.
\end{remark}

\subsection[The electromagnetic interaction]{The electromagnetic interaction}\label{EM4}\index{electromagnetic!interaction}

\noindent As mentioned above, the requirement Eq. \eqref{10} is sufficient but not necessary.  In general one may add a (possibly complex) \cite{P80} vector multiple of the unit matrix to the solution, Eq. \eqref{FI10}.  In this way we may generalize the $\Gamma_\mu{}$'s for the case where an arbitrary electromagnetic potential\index{electromagnetic!potential} $A_\mu{}$ is present \cite{P80}, \cite{KT15}.  We simply make the replacements:
\begin{align}
\Gamma_\mu{}&\rightarrow\Gamma_\mu{}+iqA_\mu{}I\,,\label{em1a}\\\nonumber\\
D_\mu{}&\rightarrow D_\mu{}+iqA_\mu{}I\,,\label{em1b}
\end{align}

\noindent where $q$ is the charge of the particle described by $\psi$.  Thus Eq. \eqref{FI13b} is now generalized to
\begin{align}
i\bar{\gamma}^{\mu}D_\mu{}\psi-m \psi&=0\,,\label{em1c}\\
i\gamma^{C}\left(e_C{}+\frac{1}{4}\,\Gamma_{ABC}\,\gamma^A{}\gamma^B{}+iqA_C{}\right)\psi-m \psi&=0\,,\label{em1d}
\end{align}

\noindent where
\begin{equation}
\label{em1e}
\bar{\gamma}^\mu{}A_\mu{}=e_C{}^\mu{}\gamma^C{}A_\mu{}=\gamma^C{}e_C{}^\mu{}A_\mu{}=\gamma^C{}A_C{}\,.
\end{equation}

\noindent This is consistent with the so-called \textit{minimal coupling} procedure. One can easily deduce the correctness of this term by considering the Minkowski limit \textup{(}e.g., see \cite{IZ80}, pp. 64-67).  For example, in the case of the hydrogenic atom,\index{hydrogenic atom} $q=-e$, $e>0$, and in the standard notation the components $A_C{}$ of electromagnetic potential  due to the proton are
\begin{equation}
\label{em4}
A=\left(A_0{}, A_1{}, A_2{}, A_3{}\right)=\left(\frac{Ze}{r}, 0, 0, 0\right)\,,
\end{equation}

\noindent so that $iqA_0{}=-i\dfrac{Ze^{2}}{r}$.\\

\noindent Of course in curved spacetime one has to use appropriate Maxwell's equations.  We refer the reader again to \cite{P80}, Sec. VII, or \cite{LL03}.\\

\subsection[The Newman-Penrose formalism]{The Newman-Penrose formalism}\label{NP}

\noindent In this section we give a short introduction to the Newman-Penrose formalism\index{Newman-Penrose!formalism} \cite{NP62}, \cite{NP63}.  Apart from the Newman-Penrose paper, we have found useful the exposition in the following texts \cite{O95}, \cite{PK06}, and \cite{NP09}.  In addition the software package \textsc{Cartan}\index{Cartan@\textsc{Cartan}}, \cite{Cv1.8}, may be used with the N-P formalism for considerably faster calculations.  However one has to be careful because, as usual, there are differences in some definitions and conventions among these references.\\

\noindent In the Newman-Penrose formalism\index{Newman-Penrose!formalism} the calculations are done using a complex null tetrad\index{tetrad!null}.  One straightforward way to construct a complex null tetrad\index{tetrad!null} for a given metric, is to choose a set, $e_A{}$, of orthonormal tetrad vector fields (as discussed in Appendix \ref{Tet}).  These satisfy
\begin{equation}
\eta_{AB}=e_A{}^{\alpha}\,e_B{}^{\beta}\,g_{\alpha\beta}\,.\label{NP8}
\end{equation}

\noindent Then we define the complex null tetrad\index{tetrad!null} $l, n, m, \overline{m},$ below \cite{GHP73}:
\begin{align}
\lambda_{1}&=l=\frac{1}{\sqrt{2}}\left(e_0{}+e_3{}\right)\,,\label{NP1}\\\nonumber\\
\lambda_{2}&=n=\frac{1}{\sqrt{2}}\left(e_0{}-e_3{}\right)\,,\label{NP2}\\\nonumber\\
\lambda_{3}&=m=\frac{1}{\sqrt{2}}\left(e_1{}+ie_2{}\right)\,,\label{NP3}\\\nonumber\\
\lambda_{4}&=\overline{m}=\frac{1}{\sqrt{2}}\left(e_1{}-ie_2{}\right).\label{NP4}
\end{align}

\noindent \textit{Note that $\overline{m}=m^{*}$, so that, in general, we will use $A^{*}$ for the complex conjugate of $A$.}  Using Eq. \eqref{NP8} we can show that the null tetrad\index{tetrad!null} vectors $l, n, m, \overline{m},$ of Eqs. \eqref{NP1} - \eqref{NP4}, satisfy the relations
\begin{align}
&l^\mu{}l_\mu{}=n^\mu{}n_\mu{}=\overline{m}^\mu{}\overline{m}_\mu{}=m^\mu{}m_\mu{}=0\,,\label{NP5}\\\nonumber\\
&l^\mu{}m_\mu{}=n^\mu{}m_\mu{}=l^\mu{}\overline{m}_\mu{}=n^\mu{}\overline{m}_\mu{}=0\,,\label{NP6}\\\nonumber\\
&l^\mu{}n_\mu{}=+1\,,\;m^\mu{}\overline{m}_\mu{}=-1\,.\label{NP7}
\end{align}

\noindent The frame field metric\index{metric!frame field} components, $\zeta_{AB}$, for the $\lambda_A{}$ of Eqs. \eqref{NP1} - \eqref{NP4}, are given by
\begin{equation}
\label{NP8}
\zeta_{AB}=\lambda_A{}^{\alpha}\,\lambda_B{}^{\beta}\,g_{\alpha\beta}=\lambda_A{}^\alpha{}\lambda_{B\alpha}\,,
\end{equation}

\noindent where $A, B=1,2,3,4$.  We find that
\begin{equation}
\label{NP9}
\left(\zeta^{AB}\right)=\left(\zeta_{AB}\right)=\left( \begin{array}{cccc} 
0 & 1 & 0 & 0 \\
\rule{0in}{5ex}
	1 & 0 & 0 & 0\\
\rule{0in}{5ex}
	0 & 0 & 0 & -1\\
\rule{0in}{5ex}
	0 & 0 & -1 & 0\\
\end{array} \right)\,.
\end{equation}

\begin{exercise}\label{NPe2}\index{Exercises}  Use Eq. \eqref{NP8} to derive Eq. \eqref{NP9}.
\end{exercise}

\noindent We also have the relations
\begin{align}
g^{\alpha\beta}&=\zeta^{AB}\lambda_A{}^\alpha{}\lambda_B{}^\beta{}\,,\label{NP9b}\\\nonumber\\
\zeta^{CB}\zeta_{AB}&=\lambda_A{}^\alpha{}\,\zeta^{CB}\lambda_{B\alpha}=\lambda_A{}^\alpha{}\,\lambda^C{}_\alpha{}=\delta_A{}^C{}\,.\label{NP9c}
\end{align}

\noindent In order to conform to the notation in the NP formalism literature, we shall change slightly the notation for the Ricci rotation coefficients\index{Ricci rotation coefficients} given by Eq. \eqref{FI22}, and write $\Gamma_{ABC}=\gamma_{ABC}$ (with $A, B, C=1,2,3,4$ where 4 is the time label).  The collection of equations below is very important and handy:
\begin{align}
\lambda^A{}_\alpha{}&=g_{\alpha\beta}\,\zeta^{AB}\lambda_B{}^\beta{}\,,\label{NP10a}\\\nonumber\\
\gamma^D{}_{BC}&=\lambda_B{}^\beta{}\lambda_C{}^\alpha{}\,\nabla_\alpha{}\,\lambda^D{}_\beta{}\,,\label{NP10b}\\\nonumber\\
\gamma_{ABC}&=\zeta_{AD}\,\gamma^D{}_{BC}\,,\label{NP10c}
\end{align}

\noindent and the equivalents to Eqs. \eqref{FI16}, \eqref{FI18}, and \eqref{FI22},
\begin{align}
C^D{}_{BC}&=\left(\lambda^D{}_{\alpha,\,\beta}-\lambda^D{}_{\beta,\,\alpha}\right)\lambda_B{}^\alpha{}\lambda_C{}^\beta{}\,,\label{NP10d}\\\nonumber\\
C_{ABC}&=\zeta_{AD}\,C^D{}_{BC}\,,\label{NP10e}\\\nonumber\\
\gamma_{ABC}&=-\frac{1}{2}\left(C_{ABC}+C_{BCA}-C_{CAB}\right)\,.\label{NP10f}
\end{align}

\begin{remark}\label{pp}\index{Remarks} It is clear from Eqs. \eqref{NP1} - \eqref{NP4} that if the tetrad, $e_A{}$, is a Fermi tetrad\index{tetrad!Fermi}, then the null tetrad\index{tetrad!null}, $\lambda_B{}$, is parallelly transported along the chosen congruence of timelike geodesics\index{congruence!of timelike geodesics}.
\end{remark}

\begin{remark}\label{equiv}\index{Remarks} A choice of a null tetrad\index{tetrad!null}, $\lambda_B{}$, is equivalent to a choice of an orthonormal tetrad, $e_A{}$, since we can solve Eqs. \eqref{NP1} - \eqref{NP4}, to express the $e_A{}$ in terms of the $\lambda_B{}$, \textup{\cite{GHP73}}, thus,
\end{remark}
\begin{align}
e_0{}&=\frac{1}{\sqrt{2}}(l+n)\,,\label{NP10.1}\\\nonumber\\
e_1{}&=\frac{1}{\sqrt{2}}(m+\overline{m})\,,\label{NP10.2}\\\nonumber\\
e_2{}&=\frac{-i}{\sqrt{2}}(m-\overline{m})\,,\label{NP10.3}\\\nonumber\\
e_3{}&=\frac{1}{\sqrt{2}}(l-n).\label{NP10.4}
\end{align}

\begin{remark}\label{pfid}\index{Remarks} From Eq. \eqref{NP10.1} we may deduce the properties\index{tetrad!properties} of the observer frame\index{frame!observer} tetrad by finding the acceleration,
\begin{equation}
\label{NP10.5}
a=\nabla_{e_{0}}e_0{}\,.
\end{equation}

\noindent Moreover, evaluating, $\nabla_{e_{0}}e_A{}\,,\;A=0,1,2,3,$ will tell us whether the tetrad is a Fermi tetrad\index{tetrad!Fermi} or not \textup{(}see Eq. \eqref{FT5}\textup{)}.
\end{remark}

\begin{exercise}\label{NPe1}\index{Exercises} Starting with Eq. \eqref{NP9b}, show that
\begin{equation}
\label{NP10g}
g_{\alpha\beta}=n_\alpha{}l_\beta{}+l_\alpha{}n_\beta{}-\overline{m}_\alpha{}m_\beta{}-m_\alpha{}\overline{m}_\beta{}\,.
\end{equation}
\end{exercise}

\noindent The null tetrad\index{tetrad!null} $l, n, m, \overline{m},$ is not uniquely defined by Eqs. \eqref{NP1}-\eqref{NP4}.  Without changing the direction of the field $l$, we may rescale it by an arbitrary factor $A$, where $A$ is a non-vanishing real function.  Thus
\begin{equation}
\label{NP11}
l^{\prime\alpha}=Al^\alpha{}.
\end{equation}

\noindent This amounts to a reparametrization of the curves tangent to $l$.  The vectors $m$ and $\overline{m}$ may be rotated in their plane by an arbitrary angle $\phi$; moreover, their scalar products with $l$, do not change when a multiple of $l$ is added to them.  Thus $m,\overline{m}$, are defined up to the transformations
\begin{equation}
\label{NP12}
m^{\prime\alpha}=e^{i\phi}m^\alpha{}+Bl^\alpha{},
\end{equation}

\noindent where $\phi$ is a real function and $B$ is a complex function.  The remaining vector $n$, may be changed by a fixed multiple of $l$ and a fixed multiple of a fixed vector in the $m,\overline{m}$ plane, so finally, we have \cite{PK06}
\begin{equation}
\label{NP13}
n^{\prime\alpha}=\frac{1}{A}\left(n^\alpha{}+B^{*}e^{i\phi}m^\alpha{}+Be^{-i\phi}\overline{m}^\alpha{}+BB^{*}l^\alpha{}\right)\,.
\end{equation}

\noindent We can express these transformations as three classes of transformations \cite{BGJ14},
\begin{align}
&l^{\prime}=l,\,m^{\prime}=m+Bl,\,n^{\prime}=n+B^{*}m+B\overline{m}+BB^{*}l,\,\;\;(\mbox{null rotation}),\label{NP14}\\
&n^{\prime}=n,\,m^{\prime}=m+Bl,\,l^{\prime}=l+B^{*}m+B\overline{m}+BB^{*}n,\,\;\;(\mbox{null rotation}),\label{NP15}\\
&l^{\prime}=Al,\,m^{\prime}=e^{i\phi}m,\,n^{\prime}=A^{-1}n,\,\;\;(\mbox{boost and orthogonal rotation}).\label{NP16}
\end{align}

\noindent We shall use the standard notation below, to designate the null tetrad\index{tetrad!null} vectors as directional derivatives\index{directional derivatives}
\begin{equation}
\label{NP17}
\lambda_1{}=l=D,\;\;\lambda_2{}=n=\Delta,\;\;\lambda_3{}=m=\delta,\;\;\lambda_4{}=\overline{m}=\delta^{*}\,.
\end{equation}

\noindent The above is expressed very clearly in reference \cite{GHP73}.  ``The role of the (vector) covariant derivative operator $\nabla_\alpha$ is taken over in the NP formalism by four scalar operators:''
\begin{equation}
\label{NP18}
D=l^\alpha{}\nabla_\alpha{},\;\;\Delta =n^\alpha{}\nabla_{\alpha},\;\;\delta=m^\alpha{}\nabla_{\alpha},\;\;\delta^{*}=\overline{m}^\alpha{}\nabla_{\alpha}\,.
\end{equation}

\noindent Thus, e.g., for any scalar function $f$ we write
\begin{equation}
\label{NP19}
Df=l^\alpha{}f_{,\alpha},\;\;\Delta f=n^\alpha{}f_{,\alpha},\;\;\delta f=m^\alpha{}f_{,\alpha},\;\;\delta^{*}f=\overline{m}^\alpha{}f_{,\alpha}\,.
\end{equation}

\noindent We give below the twelve so-called \textit{spin coefficients}\index{spin coefficients} in terms of the Ricci rotation coefficients\index{Ricci rotation coefficients}.
\begin{align}
\kappa&=\gamma_{311},\;\;\;\rho=\gamma_{314},\;\;\;\varepsilon=\tfrac{1}{2}\left(\gamma_{211}+\gamma_{341}\right)\,,\nonumber\\\nonumber\\
\sigma&=\gamma_{313},\;\;\;\mu=\gamma_{243},\;\;\;\gamma=\tfrac{1}{2}\left(\gamma_{212}+\gamma_{342}\right)\,,\nonumber\\\label{NP20}\\
\lambda&=\gamma_{244},\;\;\;\tau=\gamma_{312},\;\;\;\alpha=\tfrac{1}{2}\left(\gamma_{214}+\gamma_{344}\right)\,,\nonumber\\\nonumber\\
\nu&=\gamma_{242},\;\;\;\pi=\gamma_{241},\;\;\;\beta=\tfrac{1}{2}\left(\gamma_{213}+\gamma_{343}\right).\nonumber
\end{align}

\begin{remark}\label{opt}\index{Remarks} One could choose the functions, $A, B, \phi$, in Eqs. \eqref{NP11}, \eqref{NP12}, and \eqref{NP13}, so as to maximize the number of vanishing the spin coefficients\index{spin coefficients}, thus possibly making the Dirac equation easier to solve.  However it may not be easy to determine what observer frame\index{frame!observer} the new tetrad corresponds to \textup{(}see Remark \textup{\ref{equiv})}.\end{remark}

\noindent We finally write the Dirac equation in the N-P formalism\index{Dirac equation!in the N-P formalism} \cite{C92}, \cite{C76}, \cite{MZ94}, \cite{Z96}.
\begin{align}
(\Delta+\mu^{*}-\gamma^{*})G_{1}-(\delta^{*}+\beta^{*}-\tau^{*})G_{2}&=i\mu_{*}F_{1}\,,\nonumber\\\nonumber\\
(D+\varepsilon^{*}-\rho^{*})G_{2}-(\delta+\pi^{*}-\alpha^{*})G_{1}&=i\mu_{*}F_{2}\,,\nonumber\\\label{NP21}\\
(D+\varepsilon-\rho)F_{1}+(\delta^{*}+\pi-\alpha)F_{2}&=i\mu_{*}G_{1}\,,\nonumber\\\nonumber\\
(\Delta+\mu-\gamma)F_{2}+(\delta+\beta-\tau)F_{1}&=i\mu_{*}G_{2}\,,\nonumber
\end{align}

\noindent where $\mu_{*}\sqrt{2}=m=\mbox{the mass of the particle}$.  Eqs. \eqref{NP21} are in the chiral representation given by Eqs. \eqref{D6}, see ref. \cite{R15}.

\section[Examples]{Examples}\label{Ex}

\subsection[Schwarzschild spacetime N-P]{Schwarzschild spacetime N-P}\label{NPS}

\noindent In this section we compare the two approaches, N-P and F-I.  Chandrasekhar uses the metric signature $(+,-,-,-)$ and so we write the Schwarzschild metric\index{metric!Schwarzschild},
\begin{equation}
ds^{2}=\left(1-\frac{2M}{r}\right)dt^{2}-\displaystyle\frac{dr^{2}}{\displaystyle
\left(1-\frac{2M}{r}\right)}-r^{2}(d\theta^{2}+\sin^{2}\theta d\phi^{2})\,.\label{NPS0}
\end{equation}

\noindent We adopt essentially Chandrasekhar's null tetrad\index{tetrad!null} cf., \cite{C92}, Eq. (281) p. 134, where $l=(l^t{},l^r{},l^\theta{},l^\phi{})$, etc.
\begin{align}
l&=\frac{1}{\sqrt{2}}\left(\frac{1}{X},\,1,\,0,\,0\right)\,,\label{NPS1}\\\nonumber\\
n&=\frac{1}{\sqrt{2}}\left(1,\,-X,\,0,\,0\right)\,,\label{NPS2}\\\nonumber\\
m&=\frac{1}{\sqrt{2}}\left(0,\,0,\,\frac{1}{r},\,\frac{i}{r\sin{\theta}}\right)\,,\label{NPS3}\\\nonumber\\
\overline{m}&=\frac{1}{\sqrt{2}}\left(0,\,0,\,\frac{1}{r},\,\frac{-\,i}{r\sin{\theta}}\right),\label{NPS4}
\end{align}

\noindent where
\begin{equation}
\label{NPS4a}
X=1-\frac{2M}{r}\,.
\end{equation}

\noindent We find the spin coefficients\index{spin coefficients} below.
\begin{align}
\alpha&=-\frac{\cot{\theta}}{2\sqrt{2}\,r}\,,\label{NPS5}\\\nonumber\\
\beta&=\frac{\cot{\theta}}{2\sqrt{2}\,r}\,,\label{NPS6}\\\nonumber\\
\gamma&=\frac{M}{\sqrt{2}\,r^{2}}\,,\label{NPS7}\\\nonumber\\
\mu&=-\frac{X}{\sqrt{2}\,r}\,,\label{NPS8}\\\nonumber\\
\rho&=-\frac{1}{\sqrt{2}\,r},\label{NPS9}
\end{align}

\noindent Now we use Eqs. \eqref{NP19} and \eqref{NP21} to write the Dirac equations\index{Dirac equation!in the N-P formalism}.  We write below only the first two of the four equations for those who wish to check their results
\begin{align}
\partial_t{}\,G_1{}-X\partial_r{}\,G_1{}+\frac{M-r}{r^{2}}\,G_1{}-\frac{1}{r}\,\partial_\theta{}\,G_2{}\,+\,&\frac{i}{r\sin{\theta}}\,\partial_\phi{}\,G_2{}\,-\nonumber\\\nonumber\\&\frac{\cot{\theta}}{2r}\,G_2{}=imF_1{}\,,\label{NPS10.1}\\\nonumber\\
\frac{1}{X}\,\partial_t{}\,G_2{}+\partial_{r}\,G_2{}+\frac{1}{r}\,G_2{}-\frac{1}{r}\,\partial_\theta{}\,G_1{}\,-\,&\frac{i}{r\sin{\theta}}\,\partial_\phi\,G_1{}-\nonumber\\\nonumber\\&\frac{\cot{\theta}}{2r}\,G_1{}=imF_2{}\,.\label{NPS10.2}
\end{align}

\noindent where the Dirac wavefunction is
\begin{equation}
\label{NPS11}
\psi=\left(\begin{array}{c}
F_{1}\\
F_{2}\\
G_{1}\\
G_{2}
\end{array}\right).
\end{equation}

\noindent We remark that the details of these calculations can be carried using the software package \textsc{Cartan}\index{Cartan@\textsc{Cartan}}.\\

\noindent  To do the calculations following the Fock-Ivanenko\index{Fock-Ivanenko} approach, we recall Remark \ref{equiv} and use Eqs. \eqref{NP10.1} - \eqref{NP10.4}, to obtain the orthogonal tetrad\index{tetrad!orthogonal} corresponding to the null tetrad\index{tetrad!null} given by Eqs. \eqref{NPS1} - \eqref{NPS4}.  We have
\begin{align}
e_0{}&=\left(\frac{1+X}{2X}\,,\frac{1-X}{2},\,0,\,0\right)\,,\label{NPS12}\\\nonumber\\
e_1{}&=\left(0,\,0,\,\frac{1}{r},\,0\right)\,,\label{NPS13}\\\nonumber\\
e_2{}&=\left(0,\,0,\,0,\,\frac{1}{r\sin{\theta}}\right)\,,\label{NPS14}\\\nonumber\\
e_3{}&=\left(\frac{1-X}{2X},\,\frac{1+X}{2},\,0,\,0\right)\,,\label{NPS15}
\end{align}

\noindent where $e_A{}=(e_A{}^t{},e_A{}^r{},e_A{}^\theta{},e_A{}^\phi{})$, and $X$ is given by Eq. \eqref{NPS4a}.  Recalling Remark \ref{pfid}, we find that $a=\nabla_{e_{0}}e_0{}\neq0$, so this is not a freely falling particle frame.  Finally the Dirac equations obtained with this approach are, of course, identical to Eqs. \eqref{NPS10.1}, \eqref{NPS10.2}, etc.\\

\subsection[Schwarzschild spacetime F-I]{Schwarzschild spacetime F-I}\label{FIS}

\noindent In this example we calculate the Fock-Ivanenko \index{Fock-Ivanenko}coefficients for the Schwarzschild metric, using the tetrad below and write the resulting Dirac equations.\index{Dirac equation}  This calculation may be found in Ryder \textup{\cite{R09}} although there are several misprints there.  We use Ryder's conventions, i.e., the metric signature below, the standard representation of the $\gamma$ matrices Eqs. \eqref{D4}, \eqref{D5}, and $\varepsilon=-1$. \\

\noindent The Schwarzschild metric\index{metric!Schwarzschild} is
\begin{equation}
ds^{2}=-\left(1-\frac{2M}{r}\right)dt^{2}+\displaystyle\frac{dr^{2}}{\displaystyle
\left(1-\frac{2M}{r}\right)}+r^{2}(d\theta^{2}+\sin^{2}\theta d\phi^{2})\,,\label{E1}
\end{equation}

\noindent and we choose the orthonormal 1-forms and corresponding vectors\index{tetrad!vectors} below which satisfy $g_{\alpha\beta}=\eta_{AB}\,e^A{}_\alpha{}\,e^B{}_\beta{}$, (see Appendix \ref{GenT}).
\begin{align}
e^{0}&=\left(1-\frac{2M}{r}\right)^{\frac{1}{2}}dt\,,&e_{0}&=\left(1-\frac{2M}{r}\right)^{-\frac{1}{2}}\partial_{t}\,,\label{E2}\\\nonumber\\
e^{1}&=\left(1-\frac{2M}{r}\right)^{-\frac{1}{2}}dr\,,&e_{1}&=\left(1-\frac{2M}{r}\right)^{\frac{1}{2}}\partial_{r}\,,\label{E3}\\\nonumber\\
e^{2}&=r\,d\theta\,,&e_{2}&=\frac{1}{r}\,\partial_{\theta}\,,\label{E4}\\\nonumber\\
e^{3}&=r\sin{\theta}\,d\phi\,,&e_{3}&=\frac{1}{r\sin\theta}\,\partial_{\phi}\,.\label{E5}
\end{align}

\noindent Equations \eqref{E2} - \eqref{E5} are the tetrad 1-forms\index{tetrad!1-forms} and vectors of an observer with $\dot{r}=0, \,\dot{\theta}=0, \,\dot{\phi}=0$.  Using Eq. \eqref{FI1} we find the nonvanishing spin connection coefficients,\index{spin connection}
\begin{align}
\omega_{10t}&=\frac{M}{r^{2}}\,,\label{E7}\\\nonumber\\
\omega_{21\theta}&=\left(1-\frac{2M}{r}\right)^{\frac{1}{2}},\label{E8}\\\nonumber\\
\omega_{31\phi}&=\sin\theta\,\left(1-\frac{2M}{r}\right)^{\frac{1}{2}},\label{E9}\\\nonumber\\
\omega_{32\phi}&=\cos\theta\,.\label{E10}
\end{align}

\noindent We now use Eqs. \eqref{FI10} and \eqref{FI12b}, to obtain the Fock-Ivanenko coefficients\index{Fock-Ivanenko coefficients} $\Gamma_{C}$.
\begin{align}
\Gamma_{0}&=\frac{M}{2r^{2}}\left(1-\frac{2M}{r}\right)^{-\frac{1}{2}}\gamma^{0}\gamma^{1}\,,\label{E11}\\\nonumber\\
\Gamma_{1}&=0\,,\label{E12}\\\nonumber\\
\Gamma_{2}&=\frac{1}{2r}\left(1-\frac{2M}{r}\right)^{\frac{1}{2}}\gamma^{1}\gamma^{2}\,,\label{E13}\\\nonumber\\
\Gamma_{3}&=\frac{1}{2r}\left(1-\frac{2M}{r}\right)^{\frac{1}{2}}\gamma^{1}\gamma^{3}
\frac{\cot\theta}{2r}\gamma^{2}\gamma^{3}.\label{E14}
\end{align}

\noindent The Dirac equation,\index{Dirac equation!in Schwarzschild spacetime} \eqref{FI12b} and \eqref{FI13}, is
\begin{align}
i\big[\gamma^0{}\left(e_0{}^t{}\,\partial_t{}+\Gamma_0{}\right)&+\gamma^1{}e_1{}^r{}\,\partial_r{}+\gamma^2{}\left(e_2{}^\theta{}\,\partial_\theta{}+\Gamma_2{}\right)\nonumber\\  &{}+\gamma^3{}\left(e_3{}^\phi{}\,\partial_\phi{}+\Gamma_3{}\right)\big]\psi-m\psi=0\,.\label{E15}
\end{align}

\noindent From Eq. \eqref{E11} we see that the term $\gamma^0{}\Gamma_0{}$ appearing in Eq. \eqref{E15} simplifies to
\begin{equation}
\label{E15a}
\gamma^0{}\Gamma_0{}=\frac{M}{2r^{2}}\left(1-\frac{2M}{r}\right)^{-\frac{1}{2}}\gamma^1{}\,,
\end{equation}

\noindent and so on.

\subsection[Nonfactorizable metric]{Nonfactorizable metric}\label{HM}

\noindent In this example we shall write the Dirac equation\index{Dirac equation!in a nonfactorizable metric}\index{metric!nonfactorizable} in the spacetime considered by Hounkonnou and Mendy \cite{HM99}
\begin{equation}
\label{E15b}
ds^{2}=-dt^{2}+a^{2}(t)\left(dx^{2}+b^{2}(x)\left[dy^{2}+c^{2}(y)dz^{2}\right]\right).
\end{equation}

\noindent We remark that the de Sitter universe metric\index{metric!de Sitter} and the usual Friedman-Lema\^{i}tre-Robertson-Walker metric\index{metric!FLRW} of standard cosmology, for each curvature parameter\index{curvature!parameter} $k$ separately, are special cases of the above form.\\

\noindent As in the example of Sec. \ref{FIS} we again choose the tetrad\index{tetrad!vectors} for an observer with $\dot{x}=0, \,\dot{y}=0, \,\dot{z}=0$.  The 1-forms\index{tetrad!1-forms} and corresponding vectors are given below.
\begin{align}
e^{0}&=dt\,,&e_{0}&=\partial_{t}\,,\label{E16}\\\nonumber\\
e^{1}&=a(t)\,dx\,,&e_{1}&=\frac{1}{a(t)}\,\partial_{x}\,,\label{E17}\\\nonumber\\
e^{2}&=a(t)b(x)\,dy\,,&e_{2}&=\frac{1}{a(t)b(x)}\,\partial_{y}\,,\label{E18}\\\nonumber\\
e^{3}&=a(t)b(x)c(y)\,dz\,,&e_{3}&=\frac{1}{a(t)b(x)c(y)}\,\partial_{z}\,.\label{E19}
\end{align}

\noindent Using Eq. \eqref{FI1} we find the nonvanishing spin connection\index{spin connection} coefficients,
\begin{align}
\omega_{10x}&=a,_{t},\;\;\;\;\;\;\;\;\omega_{20y}=b\,a,_{t},\;\;\;\;\omega_{21y}=b,_{x},\label{E20}\\\nonumber\\
\omega_{30z}&=b\,c\,a,_{t},\;\;\;\;\omega_{31z}=c\,b,_{x},\;\;\;\;\omega_{32z}=c,_{z}.\label{E21}
\end{align}

\noindent Again we use Eqs. \eqref{FI10} and \eqref{FI12b}, to obtain the the Fock-Ivanenko coefficients\index{Fock-Ivanenko coefficients} $\Gamma_{C}$.
\begin{align}
\Gamma_{0}&=0\,,\label{E22}\\\nonumber\\
\Gamma_{1}&=-\frac{a,_{t}}{2a}\,\gamma^{0}\gamma^{1}\,,\label{E23}\\\nonumber\\
\Gamma_{2}&=-\frac{a,_{t}}{2a}\,\gamma^{0}\gamma^{2}-\frac{b,_{x}}{2ab}\,\gamma^{1}\gamma^{2}\,,\label{E24}\\\nonumber\\
\Gamma_{3}&=-\frac{a,_{t}}{2a}\,\gamma^{0}\gamma^{3}-\frac{b,_{x}}{2ab}\,\gamma^{1}\gamma^{3}-\frac{c,_{y}}{2abc}\,\gamma^{2}\gamma^{3}.\label{E25}
\end{align}

\noindent Note that Hounkonnou and Mendy in ref. \cite{HM99}, define their $\Gamma_\mu{}$ with opposite sign from the one adopted here, Eq. \eqref{FI10}, and in the first part of their paper they effectively multiply their $\gamma$ matrices by $(-i)$.  Thus using Eq. \eqref{Sig11} and the relations \eqref{J2}, we write the resulting Dirac equation\index{Dirac equation!in a nonfactorizable metric} as,
\begin{align}
\gamma^C{}\left(e_C{}+\Gamma_C{}\right)\psi+m\psi=&\,\bigg[\gamma^0{}\left(\partial_t{}+\frac{3a,_t{}}{2a}\right)+\gamma^{1}\left(\frac{1}{a}\partial_x{}+\frac{b,_{x}}{ab}\right)+\nonumber
\\&\gamma^2{}\left(\frac{1}{ab}\partial_y{}+\frac{c,_y{}}{2abc}\right)+\gamma^3{}\frac{1}{abc}\,\partial_z{}\bigg]\psi+m\psi=0\,.\label{E26}
\end{align}

\noindent A further simplification is achieved if we let
\begin{equation}
\label{E27}
\psi\equiv a^{-\frac{3}{2}}\,b^{-1}\,c^{-\frac{1}{2}}\Psi.
\end{equation}

\noindent A short calculation shows that we may now rewrite the Dirac equation\index{Dirac equation!in a nonfactorizable metric} in the simplified form
\begin{equation}
\label{E28}
\left[\gamma^0{}\partial_t{}+\frac{1}{a}\gamma^1{}\partial_x{}+\frac{1}{ab}\gamma^2{}\partial_y{}+\frac{1}{abc}\gamma^3{}\partial_z{}\right]\Psi+m\Psi=0\,.
\end{equation}

\noindent At some point in ref. \cite{HM99} the authors specifically adopt the Jauch-Rohrlich representation\index{representation!Jauch-Rohrlich} of the $\gamma$ matrices discussed in Sec. \ref{J-R}.  Further clarification may be obtained by reviewing Sec. \ref{C}.

\subsection[de Sitter spacetime, Fermi coordinates]{de Sitter spacetime, Fermi coordinates}\label{dS}

\noindent In this section we shall consider the Dirac equation in the de Sitter universe\index{Dirac equation!in de Sitter universe} using \textit{exact} (global) Fermi coordinates\index{Fermi coordinates} $\left(x^{0}, x^{1}, x^{2}, x^{3}\right)$ \textit{with respect to the reference observer}\index{reference observer} $(\tau,0,0,0)$.  We shall first write the de Sitter metric in the standard coordinates used in ref. \cite{CM06}, (but we adopt a different notation from the one used there in order to avoid confusion).
\begin{equation}\label{E29a}
ds^{2}= -d(y^{0})^{2}+e^{2ay^{0}}\delta_{ij}dy^i{}dy^j{}, 
\end{equation}

\noindent A Fermi tetrad field on the set of geodesics $\gamma(\tau)=(\tau,\, y^1{}_0{},\,y^2{}_0{},\,y^3{}_0{})$, i.e., $y^0{}=\tau$ and $y^i{}=\textup{const}.$, is
\begin{equation}
\label{E29b}
\lambda_0{}=\partial_{y^{0}},\;\;\;\;\lambda_I{}=e^{-a\tau}\partial_{y^{I}}, \;\;\;I=(1,2,3).
\end{equation}

\noindent We transform the metric of Eq. \eqref{E29a} to the metric in global Fermi coordinates using the transformation derived in \cite{CM06}, \cite{KC10}, relating the $y^\mu{}$ to the Fermi $x^\mu{}$,
\begin{align}
e^{ay^{0}}&=e^{ax^{0}}\cos{(a\rho)}\label{E29c}\\\nonumber\\
y^i{}&=e^{-ax^{0}}\,\frac{\tan{(a\rho)}}{a\rho}\,x^i{},\label{E29d}
\end{align}

\noindent where $\rho=\sqrt{\delta_{ij}x^i{}x^j{}}$, $a = \sqrt{ \Lambda/3}\,,$ and $0\leq\rho < \pi/(2a)$.\\

\noindent The resulting metric is,
\begin{equation}
\label{E29e}
ds^{2}= -\cos^{2}\left( a\rho \right) d(x^{0})^{2}+\left[\frac{x^{i}x^{j}}{\rho^{2}}+\frac{\sin^{2}(a\rho)}{a^{2}\rho^{2}}\left(\delta_{ij}-\frac{x^{i}x^{j}}{\rho^{2}}\right)\right]dx^{i}dx^{j},
\end{equation}

\noindent We may obtain a set of Fermi tetrad 1-form field\index{tetrad!1-form field} in Fermi coordinates by transforming the 1-forms corresponding to the vectors of Eq. \eqref{E29b}, using the transformation Eqs. \eqref{E29c}, \eqref{E29d}.  The Fermi tetrad field \index{tetrad!Fermi} obtained is complicated because the set of geodesics corresponding to the set $\gamma(\tau)$ has lost its original simplicity in the $x^\mu{}$ coordinates.  We can find the inverse of the transformation Eqs. \eqref{E29c}, \eqref{E29d}, which we shall refer to as $F$, thus
\begin{align}
x^0{}&=\displaystyle{\left(\frac{1}{a}\right)\ln{\left(\frac{e^{ay^{0}}}{\sqrt{1-a^{2}e^{2ay^{0}}R^{2}}}\right)}},\label{E29f}\\\nonumber\\
x^i{}&=\displaystyle{\frac{\arccos{\sqrt{1-a^{2}e^{2ay^{0}}R^{2}}}}{aR}\,y^i{}},\label{E29g}
\end{align}

\noindent where $R=\sqrt{\delta_{ij}y^i{}y^j{}}$.  We know that if $F:M\rightarrow N$\ is an isometry\index{isometry} and $\gamma$ is a geodesic in $M$, then $F\circ\gamma$ is a geodesic in $N$.  Therefore the set of geodesics, $\gamma(\tau)$, are now given by (recall that $y^0{}=\tau$)
\begin{equation}
\label{E29h}
F\circ\gamma=\left(x^0{}(\tau),\,x^1{}(\tau),\,x^2{}(\tau),\,x^3{}(\tau)\right).
\end{equation}

\noindent The above complications do not prevent us from carrying on with our calculations for the Dirac equation.  We give the set of tetrad 1-forms\index{tetrad!1-forms} we obtained below.\\
\begin{align}
e^{0}&=dx^{0}-f(\rho)\left(x^{1}\,dx^{1}+x^{2}\,dx^{2}+x^{3}\,dx^{3}\right)\,,\label{E30}\\\nonumber\\
e^{1}&=-x^{1}h(\rho)\,dx^{0}\nonumber\\\nonumber\\&+\frac{a\rho\left(x^{1}\right)^{2}\sec{(a\rho)}+\left[\left(x^{2}\right)^{2}+\left(x^{3}\right)^{2}\right]\sin{(a\rho)}}{a\rho^{3}}\,dx^{1}\nonumber\\\nonumber\\ &+x^{1}x^{2}\,p(\rho)\,dx^{2}+x^{1}x^{3}\,p(\rho)\,dx^{3},\label{E31}\\\nonumber\\
e^{2}&=-x^{2}\,h(\rho)\,dx^{0}+x^{1}x^{2}\,p(\rho)\,dx^{1}\nonumber\\\nonumber\\&+\frac{a\rho\left(x^{2}\right)^{2}\sec{(a\rho)}+\left[\left(x^{1}\right)^{2}+\left(x^{3}\right)^{2}\right]\sin{(a\rho)}}{a\rho^{3}}\,dx^{2}\nonumber\\\nonumber\\&+x^{2}x^{3}\,p(\rho)\,dx^{3},\label{E32}\\\nonumber\\
e^{3}&=-x^{3}\,h(\rho)\,dx^{0}+x^{1}x^{3}\,p(\rho)\,dx^{1}+x^{2}x^{3}\,p(\rho)\,dx^{2}\nonumber\\\nonumber\\&+\frac{a\rho\left(x^{3}\right)^{2}\sec{(a\rho)}+\left[\left(x^{1}\right)^{2}+\left(x^{2}\right)^{2}\right]\sin{(a\rho)}}{a\rho^{3}}\,dx^{3},\label{E33}
\end{align}

\noindent where
\begin{align}
f(\rho)&=\frac{\tan{(a\rho)}}{\rho}\,,\label{E30d}\\\nonumber\\
h(\rho)&=\frac{\sin{(a\rho)}}{\rho},\label{E31d}\\\nonumber\\
p(\rho)&=\left(\frac{\sec{(a\rho)}}{\rho^{2}}-\frac{\sin{(a\rho)}}{a\rho^{3}}\right),\label{E32d}\\\nonumber\\
q(\rho)&=\left(-\frac{a\csc{(a\rho)}}{\rho}+\frac{\sec{(a\rho)}}{\rho^{2}}\right).\label{E33d}
\end{align}

\noindent  The corresponding tetrad vectors\index{tetrad!vectors} are,\\
\begin{align}
e_{0}&=\sec^{2}{(a\rho)}\,\partial_{x^{0}}+f(\rho)\left(x^{1}\,\partial_{x^{1}}+x^{2}\,\partial_{x^{2}}+x^{3}\,\partial_{x^{3}}\right),\label{E30v}\\\nonumber\\
e_{1}&=x^{1}f(\rho)\sec{(a\rho)}\,\partial_{x^{0}}\nonumber\\\nonumber\\ &+\frac{a\rho\left[\left(x^{2}\right)^{2}+\left(x^{3}\right)^{2}\right]\csc{(a\rho)}+\left(x^{1}\right)^{2}\sec(a\rho)}{\rho^{2}}\,\partial_{x^{1}}\nonumber\\\nonumber\\ &+q(\rho)\left(x^{1}x^{2}\,\partial_{x^{2}}+x^{1}x^{3}\partial_{x^{3}}\right),\label{E31v}\\\nonumber\\
e_{2}&=x^{2}f(\rho)\sec{(a\rho)}\,\partial_{x^{0}}+x^{1}x^{2}\,q(\rho)\,\partial_{x^{1}}\nonumber\\\nonumber\\&+\frac{a\rho\left[\left(x^{1}\right)^{2}+\left(x^{3}\right)^{2}\right]\csc{(a\rho)}+\left(x^{2}\right)^{2}\sec(a\rho)}{\rho^{2}}\,\partial_{x^{2}}\nonumber\\\nonumber\\ &+x^{2}x^{3}\,q(\rho)\,\partial_{x^{3}},\label{E32v}\\\nonumber\\
e_{3}&=x^{3}f(\rho)\sec{(a\rho)}\,\partial_{x^{0}}+q(\rho)\left(x^{1}x^{3}\,\partial_{x^{1}}+x^{2}x^{3}\partial_{x^{2}}\right)\nonumber\\\nonumber\\&+\frac{a\rho\left[\left(x^{1}\right)^{2}+\left(x^{2}\right)^{2}\right]\csc{(a\rho)}+\left(x^{3}\right)^{2}\sec(a\rho)}{\rho^{2}}\,\partial_{x^{3}}.\label{E33v}
\end{align}

\noindent We now use the Mathematica package \textsc{Cartan}\index{Cartan@\textsc{Cartan}} \cite{Cv1.8} to obtain the nonvanishing Ricci rotation coefficients, $\Gamma_{ABC}$, of Eqs. \eqref{FI21}, \eqref{FI22}, (but see comment below \eqref{FI22}).  Then the Fock-Ivanenko coefficients,\index{Fock-Ivanenko coefficients} $\Gamma_C{}$, of Eq. \eqref{FI24}.  We have
\begin{equation}
\label{dS1}
\Gamma_{101}=\Gamma_{202}=\Gamma_{303}=a,
\end{equation}

\noindent and
\begin{equation}
\label{dS2}
\Gamma_{0}=0,\;\;\;\;\Gamma_{A}=-\frac{a}{2}\,\gamma^{0}\gamma^{A},\;\;\;A=1,2,3.
\end{equation}

\noindent Using Eqs. \eqref{dS2}, we write the Dirac equation\index{Dirac equation!in de Sitter universe} \eqref{Sig9},
\begin{align}
\bigg[\gamma^0{}\,e_0{}&+\gamma^1{}\left(e_1{}-\frac{a}{2}\gamma^0{}\,\gamma^1{}\right)+\gamma^2{}\left(e_2{}-\frac{a}{2}\gamma^0{}\,\gamma^2{}\right)+\nonumber
\\&\gamma^3{}\left(e_3{}-\frac{a}{2}\gamma^0{}\,\gamma^3{}\right)\bigg]\psi-m\psi=0\,.\label{dS3}
\end{align}

\noindent We note that because of our $(-,+,+,+,)$ signature, the $\gamma$ matrices will satisfy Eq. \eqref{J2} and the Dirac equation,\index{Dirac equation!in de Sitter universe} simplifies to
\begin{equation}
\label{dS4}
\bigg[\gamma^A{}\,e_A{}+\gamma^0{}\left(e_0{}+\frac{3a}{2}\right)\bigg]\psi-m\psi=0,\;\;\;A=1,2,3.
\end{equation}

\noindent Equation \eqref{dS4} looks deceptively simple, but the complications arise from the expressions for the $e_A{}$, therefore it may be better to work in the original coordinates, Eq. \eqref{E29a} with the original tetrad Eq. \eqref{E29b}.

\section[The Dirac equation in (1+1) GR]{The Dirac equation in (1+1) GR}\label{1Dir}

\subsection[Introduction to (1+1)]{Introduction to (1+1)}\label{Intro1}

\noindent We shall adopt the metric signature $(+,-)$.  In (1+1) general relativity the Dirac equation simplifies and may be written as follows \cite{MMSS91}, \cite{SR94}, \cite{MM91}.
\begin{equation}
\label{1D1}
\left[i\gamma^A{}e_A{}^\mu{}\partial_\mu{}+\frac{i}{2}\gamma^A{}\frac{1}{\sqrt{-g}}\partial_\mu{}\left(\sqrt{-g}\,e_A{}^\mu{}\right)-mI_{2}\right]\psi=0\,,
\end{equation}

\noindent where the zweibein\index{zweibein} vector label $A$ runs over $0,1$, and for the spinor we write
\begin{equation}
\label{1D2}
\psi=\left(\begin{array}{c}\psi_{1}\\
\psi_{2}
\end{array}\right).
\end{equation}

\noindent In what follows we will further restrict ourselves to the \textit{chiral} (Weyl) representation of the Dirac $\gamma$ matrices, specifically we choose \cite{MMSS91}
\begin{equation}
\label{1D3}
\gamma^{0}=\left(
 \begin{matrix}
     0 & 1\\
     \rule{0in}{2ex}
     1 & 0\\
 \end{matrix}
 \right),\;\;\;\;\gamma^{1}=\left(
 \begin{matrix}
     0 &  -1\\
     \rule{0in}{2ex}
    1 &\; 0\\
 \end{matrix}
 \right).
\end{equation}

\noindent Thus
\begin{equation}
\label{1D4}
\left(\gamma^{0}\right)^{2}=I_{2}\,,\hspace{0.5cm}\left(\gamma^{1}\right)^{2}=-I_{2}\,,
\end{equation}

\noindent and
\begin{equation}
\label{1D5}
\{\gamma^{A},\gamma^{B}\}=2\eta^{AB}I_{2}\,.
\end{equation}

\noindent We also define the matrix
\begin{equation}
\label{1D6}
\gamma^{5}:=\gamma^{0}\gamma^{1}
=\left(
 \begin{matrix}
     1 & 0\\
     \rule{0in}{2ex}
     0 & -1\\
 \end{matrix}
 \right).
\end{equation}

\noindent One great advantage of the chiral representation is the ease of decoupling of Eq. \eqref{1D1}.  Of course the spinor wave function components of Eq. \eqref{1D2} are now eigenstates of the operator $\gamma^{5}$, so we may write
\begin{equation}
\label{1D7}
\psi=\left(\begin{array}{c}\psi_{(+)}\\
\psi_{(-)}
\end{array}\right),
\end{equation}

\noindent with the eigenvalues $\gamma^5{}\psi_{(+)}=+\,\psi_{(+)}$, and, $\gamma^5{}\psi_{(-)}=-\,\psi_{(-)}$.

\subsection[The Dirac equation in the Milne universe]{The Dirac equation in the Milne universe}\label{1DM}

\noindent We shall consider solutions of the Dirac equation in the Milne universe\index{Milne universe}\index{universe!Milne} in two different  charts:  (a) in standard comoving coordinates\index{comoving coordinates} $(t,x)$, in which case the metric is
\begin{equation}
\label{1E1}
ds^2{}=dt^2{}-a_{0}^{2}\,t^2{}dx^2{}\,,
\end{equation}

\noindent and (b) in exact Fermi coordinates\index{Fermi coordinates} $(\tau,\rho)$, in which case the Milne universe\index{Milne universe}\index{universe!Milne} is the interior of the forward lightcone of Minkowski spacetime\index{Minkowski}, \cite{KR11}, \cite{KR18} thus
\begin{equation}
\label{1E2}
ds^2{}=d\tau^2{}-d\rho^2{}\,,
\end{equation}

\noindent where $\tau > |\rho |$.\\

\noindent \textbf{(a)} We use the default zweibein\index{zweibein}
\begin{equation}
\label{1E3}
\bar{e}_{0}=\partial_{t},\;\;\;\;\;\bar{e}_{1}=\frac{1}{a_{0}\,t}\,\partial_{x}\,.
\end{equation}

\noindent Since the metric in Eq. \eqref{1E1} does not depend on $x$ the corresponding canonical momentum $p_{x}$ is a constant both in classical and quantum mechanics.  We take advantage of this fact and write the 2-component spinor $\psi$ as
\begin{equation}
\label{1E4}
\psi(t,x)=\left(\begin{array}{c}\psi_{1}\\
\psi_{2}
\end{array}\right)=e^{-ip_x{}x}
\left(\begin{array}{c}f_1{}(t)\\
f_2{}(t)
\end{array}\right),
\end{equation}

\noindent where $p_{x}$ is the 1-form of the particle's momentum.\\

\noindent One finds that the only nonvanishing term from the second set of terms in Eq. \eqref{1D1} is
\begin{equation}
\label{1E5}
\frac{i}{2}\gamma^0{}\frac{1}{a_{0}t}\left[\partial_{t}\left(a_{0}t\,\bar{e}_0{}^t{}\right)\right]=\frac{i}{2t}\gamma^0{}\,.
\end{equation}

\noindent Thus Eq. \eqref{1D1} reduces to
\begin{equation}
\label{1E6}
\left[i\gamma^0{}\partial_{t}+\frac{i}{a_{0}t}\,\gamma^{1}\partial_{x}+\frac{i}{2t}\gamma^0{}-mI_{2}\right]\psi=0\,.
\end{equation}

\noindent Now we substitute Eqs. \eqref{1D3} and \eqref{1E4} in Eq. \eqref{1E6} and obtain the coupled equations,
\begin{align}
f_{1}&=\frac{1}{m}\left(i\partial_{t}-\frac{p_x{}}{a_{0}t}+\frac{i}{2t}\right)f_{2}\,,\label{1E7}\\\nonumber\\
f_{2}&=\frac{1}{m}\left(i\partial_{t}+\frac{p_x{}}{a_{0}t}+\frac{i}{2t}\right)f_{1}\,.\label{1E8}
\end{align}

\noindent Finally, decoupling Eqs. \eqref{1E7} and \eqref{1E8}, we obtain
\begin{equation}
\label{1E9}
t^{2}f_{1}^{\prime\prime}+tf_{1}^{\prime}+\left[m^{2}\,t^{2}-\left(\frac{1}{2}-\frac{ip_x{}}{a_{0}}\right)^{\!2}\,\right]\,f_{1}=0,
\end{equation}

\noindent where the primes denote derivatives with respect to $t$.  The solution of this equation is given below in terms of the Bessel functions $J_{\nu}$ and $Y_{\nu}$ of the first and second kind respectively,
\begin{equation}
\label{1E10}
f_{1}(t)=AJ_{\nu}(mt)+BY_{\nu}(mt)\,,
\end{equation}

\noindent where $A$ and $B$ are arbitrary (complex) constants and 
\begin{equation}
\label{1E12}
\nu=\frac{1}{2}-\frac{ip_x{}}{a_{0}}\,.
\end{equation}

\noindent  Using Eq. \eqref{1E8}, we find that
\begin{equation}
\label{1E11}
f_{2}(t)=iAJ_{\nu-1}(mt)+iBY_{\nu-1}(mt)\,.
\end{equation}

\noindent \textbf{(b)} Now we transform the solution to the exact Fermi coordinates\index{Fermi coordinates} $(\tau,\rho)$.  The transformation and its inverse is given by \cite{KR11}
\begin{align}
t&=\sqrt{\tau^{2}-\rho^{2}}\,,\label{1E13}\\\nonumber\\
x&=\left(\frac{1}{a_{0}}\right)\tanh^{-1}{\left(\frac{\rho}{\tau}\right)}\,,\label{1E14}\\\nonumber\\
\tau&=t\cosh{(a_{0}x)}\,,\label{1E15}\\\nonumber\\
\rho&=t\sinh{(a_{0}x)}\,.\label{1E16}
\end{align}

\noindent Under the above coordinate transformation the original tetrad 1-form fields,
\begin{equation}
\label{1E17}
\bar{e}^{0}=dt,\;\;\;\;\;\bar{e}^{1}=a_{0}\,t\,dx\,,
\end{equation}

\noindent transform into the 1-form fields $h^A{}$ below
\begin{align}
h^0{}&=\frac{\tau}{\sqrt{\tau^{2}-\rho^{2}}}\,d\tau-\frac{\rho}{\sqrt{\tau^{2}-\rho^{2}}}\,d\rho,\label{1E18}\\\nonumber\\
h^1{}&=\frac{-\rho}{\sqrt{\tau^{2}-\rho^{2}}}\,d\tau+\frac{\tau}{\sqrt{\tau^{2}-\rho^{2}}}\,d\rho,\label{1E19}
\end{align}

\noindent which, of course, satisfy the relation
\begin{equation}
\label{1E20}
h^A{}_\alpha{}h^B{}_\beta{}\,\eta^{\alpha\beta}=\eta^{AB},
\end{equation}

\noindent where the upper case latin indices run over $0,1$, while the greek indices run over $\tau,\rho$.  We shall write $\psi_{h}(\tau,\rho)$ for the $\psi(t,x)$ of Eq. \eqref{1E4} transformed using Eqs. \eqref{1E13} and \eqref{1E14}.  Thus
\begin{equation}
\label{1E21}
\psi_{h}(\tau,\rho)=e^{-ip_x{}\left(\frac{1}{a_{0}}\right)\tanh^{-1}{\left(\frac{\rho}{\tau}\right)}}
\left(\begin{array}{c}f_1{}(\sqrt{\tau^{2}-\rho^{2}})\\
f_2{}(\sqrt{\tau^{2}-\rho^{2}})
\end{array}\right)\,.
\end{equation}

\noindent We will now perform a local Lorentz transformation,\index{Lorentz!tranformation} $\Lambda$, which will transform the zweibein 1-form fields $h^A{}$ into the canonical zweibein\index{canonical zweibein} for the metric \eqref{1E2},
\begin{equation}
\label{1E22}
e^0{}=d\tau,\;\;\;\;\;e^1{}=d\rho\,.
\end{equation}

\noindent The transformation $\Lambda$ is given by
\begin{equation}
\label{1E23}
e^A{}_\alpha{}=\Lambda^A{}_B{}\,h^B{}_\alpha{}\,.
\end{equation}

\noindent We find
\begin{equation}
\label{1E24}
\Lambda
=\left(
 \begin{matrix}
     \Lambda^0{}_0{} & \Lambda^0{}_1{}\\
     \rule{0in}{2ex}
     \Lambda^1{}_0{} & \Lambda^1{}_1{}\\
 \end{matrix}
 \right)=\frac{1}{\sqrt{\tau^{2}-\rho^{2}}}\left(
 \begin{matrix}
     \tau & \rho\\
     \rule{0in}{2ex}
     \rho & \tau\\
 \end{matrix}\right).
\end{equation}

\noindent We can calculate the matrix $L$ used in Eq. \eqref{1a} following the prescription given in Appendix $B$ of \cite{CK18},
\begin{equation}
\label{1E25}
L
=\left(
 \begin{matrix}
     \left(\frac{\tau+\rho}{\tau-\rho}\right)^{1/4} & 0\\
     \rule{0in}{2ex}
     0 &  \left(\frac{\tau-\rho}{\tau+\rho}\right)^{1/4}\\
 \end{matrix}
 \right).
\end{equation}

\noindent The solutions using the zweibein\index{zweibein} set $e_{A}$ Eq. \eqref{1E22} is
\begin{equation}
\label{1E26}
\psi_{e}(\tau,\rho)=L\,\psi_{h}(\tau,\rho),
\end{equation}

\noindent where $\psi_{h}(\tau,\rho)$ is given by Eq. \eqref{1E21}.  We have then,
\begin{equation}
\label{1E27}
\psi_{e}(\tau,\rho)=e^{-ip_x{}\tanh^{-1}{\left(\frac{\rho}{\tau}\right)}}
\left(\begin{array}{c}\left(\frac{\tau+\rho}{\tau-\rho}\right)^{1/4}\left(A\,J_{\nu}(z)+B\,Y_{\nu}(z)\right)\\\\
i\left(\frac{\tau-\rho}{\tau+\rho}\right)^{1/4}\left(A\,J_{\nu-1}(z)+B\,Y_{\nu-1}(z)\right)
\end{array}\right)\,,
\end{equation}

\noindent where
\begin{align}
\nu&=\frac{1}{2}-\frac{ip_x{}}{a_{0}}\,,\label{1E28}\\\nonumber\\
z&=m\sqrt{\tau^{2}-\rho^{2}}\,.\label{1E29}
\end{align}

\noindent The wavefunction $\psi_{e}(\tau,\rho)$, satisfies Eq. \eqref{1D1} which now reduces to the usual Minkowski spacetime Dirac equation, namely,
\begin{equation}
\label{1E30}
\left(i\gamma^A{}\partial_{A}-mI_{2}\right)\psi_{e}=0.
\end{equation}

\noindent In order to show that $\psi_{e}(\tau,\rho)$ satisfies Eq. \eqref{1E30}, one has to use the Bessel function\index{Bessel functions} identity,
\begin{equation}
\label{1E31}
C_{\nu-1}(z)+C_{\nu+1}(z)=\frac{2\nu}{z}C_{\nu}(z)\,,
\end{equation}

\noindent where $C_{\nu}(z)$ denotes either of the Bessel functions\index{Bessel functions} $J_{\nu}(z)\,,Y_{\nu}(z)$.\\

\noindent \textbf{(c)} In this subsection we shall consider the normalization integral\index{normalization integral} \cite{IZ80}, p. 69.  This integral is referred to as the ``probability integral''\index{probability integral} in \cite{FR09}.  Thus in Fermi coordinates\index{Fermi coordinates} with the canonical zweibein\index{canonical zweibein} we have,
\begin{equation}
\label{1E32}
\left(\psi_{e}\,\vert\,\psi_{e}\right)=\int_{\Sigma}\bar{\psi_{e}}\,\gamma^0{}\psi_{e}\, d\rho\,.
\end{equation}

\noindent In our case the spacelike hypersurface\index{hypersurface!spacelike} $\Sigma$ is the usual $(\tau_{0},\rho),\;\tau_{0}>0,$ hyperplane.  Using Eq. \eqref{1D4} we have
\begin{equation}
\label{1E33}
\left(\psi_{e}\,\vert\,\psi_{e}\right)=\int_{-\tau_{0}}^{\tau_{0}}\psi_{e}^{\dagger}\,\psi_{e}\, d\rho\,.
\end{equation}

\noindent For the remainder of this section we will write $\bar{a}$ to denote the complex conjugate of $a$.  We shall make use of the fact that
\begin{equation}
\label{1E34}
\overline{C_{\nu}(z)}=C_{\bar{\nu}}(\bar{z}),
\end{equation}

\noindent where again $C_{\nu}(z)$ denotes either of the Bessel functions\index{Bessel functions} $J_{\nu}(z)\,,Y_{\nu}(z)$.  Since in our case $z$ is real we have that
\begin{equation}
\label{1E35}
\overline{C_{\nu}(z)}=C_{\bar{\nu}}(z).
\end{equation}

\noindent In order to check the behavior of the integrand in Eq. \eqref{1E33} at the endpoints of integration, we use the limiting forms of the Bessel functions when $\nu$ is fixed and $z\sim 0$.  Using Abramowitz and Stegun's Eqs. (9.1.2), p.358 and (9.1.7), p.360, \cite{AS72}, one easily deduces that for $\nu\neq$ negative integer,
\begin{align}
Y_{\nu}(z)&=\frac{J_{\nu}(z)\cos{(\nu\pi)}-J_{-\nu}(z)}{\sin{(\nu\pi)}},\label{1E36}\\\nonumber\\
J_{\nu}(z)&\sim\left(\frac{z}{2}\right)^{\nu}\frac{1}{\Gamma(\nu+1)},\label{1E37}\\\nonumber\\
Y_{\nu}(z)&\sim-\left(\frac{z}{2}\right)^{-\nu}\frac{\Gamma(\nu)}{\pi},\;\;\;\;\;\;\operatorname{Re}(\nu)>0,\label{1E38}\\\nonumber\\
Y_{\nu}(z)&\sim-\left(\frac{z}{2}\right)^{\nu}\frac{\cot{(\nu\pi)}}{\Gamma(\nu+1)},\;\;\operatorname{Re}(\nu)<0.\label{1E39}
\end{align}

\noindent Using Eqs. \eqref{1E37}-\eqref{1E39} in the integrand of Eq. \eqref{1E33}, we see that some terms blow up at the endpoints like $\sim 1/(\tau_{0}-\rho)$ as the integration variable $\rho\rightarrow\tau_{0}$, regardless of the value of $p_x{}$.  We can eliminate these terms by setting $B=B_{1}+iB_{2}=0$, where $B_{1}, B_{2}\in \text{Reals}$.  Thus the solution $\psi_{e}$ of Eq. \eqref{1E27} reduces to
\begin{equation}
\label{1E40}
\psi_{e}(\tau,\rho)=A\,e^{-ip_x{}\tanh^{-1}{\left(\frac{\rho}{\tau}\right)}}
\left(\begin{array}{c}\left(\frac{\tau+\rho}{\tau-\rho}\right)^{1/4}\,J_{\nu}(z)\\\\
i\left(\frac{\tau-\rho}{\tau+\rho}\right)^{1/4}\,J_{\nu-1}(z)\
\end{array}\right)\,,
\end{equation}

\noindent where $A$ is a complex normalization constant.

\section[Scalar product]{Scalar product}\label{SP}

\subsection[Conservation of $j$ in SR]{Conservation of $j$ in SR}\label{conj}

\noindent The probability current density\index{probability current!density} for a Dirac field is given by
\begin{equation}
\label{N1}
j^A{}=\bar{\psi}\gamma^A{}\psi\,,
\end{equation}

\noindent where the adjoint spinor\index{spinor!adjoint} is $\bar{\psi}=\psi^{\dagger}\gamma^{0}$.  The probability current density $j^A{}$ transforms like a 4-vector under a Lorentz transformation, $\Lambda$, so that,
\begin{equation}
\label{N1a}
j^{\prime A}=\Lambda^A{}_B{}\,j^B{},
\end{equation}

\noindent moreover $j^A{}$ is conserved, that is,
\begin{equation}
\label{N2}
\partial_A{}j^A{}=0\,.
\end{equation}

\noindent First we will prove Eq. \eqref{N1a}.  We know that under a tetrad rotation (local Lorentz transformation) $\Lambda$,
\begin{equation}
\label{N1b}
\psi^{\prime}=L\psi\;\;\;\;\Rightarrow\;\;\;\;\psi^{\prime\dagger}=\psi^{\dagger}L^{\dagger}\,,
\end{equation}

\noindent where $L$ and $\Lambda$ are related as in \eqref{1a}.  Thus, using Eqs. \eqref{1S7} and \eqref{1S8}, we have that
\begin{align}
j^{\prime A}&=\psi^{\prime\dagger}\gamma^{0}\gamma^{A}\psi^{\prime},\label{N1c}\\
&=\psi^{\dagger}L^{\dagger}\gamma^{0}\gamma^{A}L\psi\,\label{N1d}\\
&=\psi^{\dagger}\gamma^{0}\left(L^{-1}\gamma^{A}L\right)\psi,\label{N1e}\\
&=\psi^{\dagger}\gamma^{0}\Lambda^A{}_B{}\gamma^{B}\psi,\label{N1f}\\
&=\Lambda^A{}_B{}\psi^{\dagger}\gamma^{0}\gamma^{B}\psi,\label{N1g}\\
&=\Lambda^A{}_B{}\,j^B{}\,.\label{N1h}
\end{align}

\noindent Next in order to show that the current is conserved it will be useful to write the expression for the adjoint of the Dirac equation\index{Dirac equation!adjoint}.  We begin with the usual special relativity Dirac equation,
\begin{equation}
\label{N1i}
i\gamma^A{}\partial_{A}\psi-m\psi=i\slashed{\partial}\,\psi-m\psi=0.
\end{equation}

\noindent Then the adjoint is obtained as follows:
\begin{align}
\left(i\gamma^A{}\partial_{A}\psi-m\psi\right)^{\dagger}&=0\,,\label{N1j}\\
-i\partial_{A}\psi^{\dagger}\left(\gamma^A{}\right)^{\dagger}-m\psi^{\dagger}&=0\,,\label{N1k}\\
i\partial_{A}\psi^{\dagger}\gamma^0{}\gamma^A{}\gamma^0{}+m\psi^{\dagger}&=0\,,\label{N1l}\\
i\left(\partial_{A}\bar{\psi}\right)\gamma^A{}+m\bar{\psi}&=0\,,\label{N1m}
\end{align}

\noindent where we used the relations \eqref{D3} and \eqref{D7}.  Thus we shall use the notation
\begin{align}
\bar{\psi}\overleftarrow{\slashed{\partial}}&=\partial_{A}\bar{\psi}\gamma^A{}\,,\label{N2a}\\
\overrightarrow{\slashed{\partial}}\psi&=\partial_{A}\psi\gamma^A{}\,.\label{N2b}
\end{align}

\noindent So we have the shorthand Feynman slash notation\index{Feynman!slash notation} for the Dirac equation and its adjoint:
\begin{align}
\left(i\overrightarrow{\slashed{\partial}}-mI\right)\psi&=0\,,\label{N3}\\
\bar{\psi}\left(i\overleftarrow{\slashed{\partial}}+mI\right)&=0\,.\label{N4}
\end{align}

\noindent Multiplying Eq. \eqref{N3} on the left with $\bar{\psi}$, and Eq. \eqref{N4} on the right with $\psi$ and adding, we obtain
\begin{equation}
\label{N6}
\bar{\psi}\left(\overleftarrow{\slashed{\partial}}+\overrightarrow{\slashed{\partial}}\right)\psi\equiv\partial_A{}\left(\bar{\psi}\gamma^A{}\psi\right)=0,
\end{equation}

\noindent which completes the proof of the conservation Eq. \eqref{N2}.\\

\noindent It follows from Eq. \eqref{N2} that
\begin{equation}
\label{CJ3}
\frac{d}{dt}\int_{V}j^0{}d^{3}x=-\int_{V}\partial_K{}\,j^K{} d^{3}x=-\int_{\partial V}j^K{}dS_K{}=0,\;\;\;K=(1,2,3).
\end{equation}

\noindent In Eq. \eqref{CJ3} we have used Gauss' theorem\index{Gauss' theorem} where $\partial V$ is the boundary of the volume $V$, so that we may write $j^K{}dS_K{}=j^1{}\,dx^2{}\wedge dx^3{}+j^2{}\,dx^3{}\wedge dx^1{}+j^3{}\,dx^1{}\wedge dx^2{}.$  The last step of Eq. \eqref{CJ3} is valid for infinite volumes, with surface at infinity, provided $\psi$ vanishes sufficiently fast there.  Eq. \eqref{CJ3} \textit{is an expression of conservation of \textup{(}total\textup{)} probability in time.}

\subsection[The current density in GR]{The current density in GR}\label{JGR}

\noindent The probability current density in general relativity (curved spacetime) is given by
\begin{equation}
\label{GR1}
j^\alpha{}=\bar{\psi}\,\bar{\gamma}^\alpha{}(x)\psi,
\end{equation}

\noindent where $\psi$ is a solution of Eq. \eqref{FI13a}, the $\bar{\gamma}^{\alpha}(x)$ are given by Eq. \eqref{FI1b} and $\bar{\psi}=\psi^{\dagger}\gamma^{0}$.  The curved spacetime proof of Eq. \eqref{N2} is given in \cite{P80} and \cite{PT09}, p. 145, and follows similar steps as the above derivation except that partial derivatives become covariant derivatives and, of course, one has to use Eq. \eqref{FI13a} instead of the Minkowski spacetime Dirac equation.  A discussion of the generalization of Eq. \eqref{CJ3} is given in Appendix E in ref. \cite{C04}.  We also refer the reader to our Proposition \ref{conserv2} below.

\subsection[The Scalar product in SR]{The Scalar product in SR}\label{SPSR}

\noindent We now define the usual scalar product\index{scalar product!in SR} (see for example ref. \cite{IZ80}, p. 69)
\begin{equation}
\label{SP1}
\left(\phi\,\vert\,\psi\right)=\int_{\Sigma}\bar{\phi}\,\gamma^0{}\psi\, d^{3}x\,.
\end{equation}

\noindent In the case where our spacelike hypersurface\index{hypersurface!spacelike} $\Sigma$ is not the usual $(t_{0},x,y,z)$ hyperplane, Eq. \eqref{SP1} generalizes to
\begin{equation}
\label{SP2}
\left(\phi\,\vert\,\psi\right)=\int_{\Sigma}\bar{\phi}\,\gamma^A{}n_A{}\,\psi\,d\Sigma\,,
\end{equation}

\noindent where $n$ is the future-directed normal to $\Sigma$, and $d\Sigma$ is the invariant ``volume element'' on $\Sigma$.  The probability integral\index{probability integral} $(\psi\vert\psi)$ is then given by
\begin{equation}
\label{SP3}
\left(\psi\,\vert\,\psi\right)=\int_{\Sigma} j^A{}n_A{}\,d\Sigma\,.
\end{equation}

\noindent In special relativity one usually chooses $\Sigma$ to be the $t=0$ hyperplane.

\subsection[Scalar product in GR]{Scalar product in GR}\label{SPGR}

\noindent We follow ref. \cite{FR09} and define the scalar product\index{scalar product!in GR}
\begin{equation}
\label{SP4}
\left(\phi\,\vert\,\psi\right)=\int_{\Sigma}\bar{\phi}\,\bar{\gamma}^\alpha{}(x)n_\alpha{}\psi\, d\Sigma\,,
\end{equation}

\noindent where $\bar{\phi}=\phi^{\dagger}\gamma^{0}$, and the $\bar{\gamma}^\alpha{}(x)$ are given by Eq. \eqref{FI1b}.  The vector $n$ is the future-directed normal to the spacelike Cauchy\index{hypersurface!Cauchy} hypersurface\index{hypersurface!spacelike} $\Sigma$, and $d\Sigma$ is the invariant ``volume element''\index{volume element} on $\Sigma$.  Using Eq. \eqref{GR1} we have that the probability integral\index{probability integral} $(\psi\vert\psi)$ is given by
\begin{equation}
\label{SP6}
\left(\psi\,\vert\,\psi\right)=\int_{\Sigma} j^\alpha{}n_\alpha{}\,d\Sigma\,.
\end{equation}

\noindent We briefly comment on Parker's definitions \cite{P80}, \cite{HP09}.  Parker defines the current\index{probability current!in GR} of Eq. \eqref{GR1} with a minus sign in front.  This is necessary because he has chosen a representation where $\left(\gamma^0{}\right)^{2}=-I$ and metric the signature $(-,+,+,+)$ (see also his argument regarding the positive definiteness of $\left(\psi\,\vert\,\psi\right)$ around his Eq. (3.5)).  In addition he defines the scalar product\index{scalar product!in GR}
\begin{equation}
\label{SP7}
\left(\phi\,\vert\,\psi\right)=-\int_{\Sigma}\bar{\phi}\,\bar{\gamma}^0{}(x)\psi\,\sqrt{-g}\,d^{3}x\,,
\end{equation}

\noindent where the integration is over a constant $x^0{}$ Cauchy hypersurface\index{hypersurface!Cauchy} ($d^{3}x=d\Sigma_t{}$).  Parker is essentially using the \textit{lapse}\index{lapse} \textit{and shift}\index{shift} formulation \cite{P04}, where the metric is written in the $(3+1)$ decomposition (in his signature)
\begin{equation}
\label{SP8}
ds^{2}=-N^{2}dt^{2}+h_{\alpha\beta}\left(dx^\alpha{}+N^\alpha{}dt\right)\left(dx^\beta{}+N^\beta{}dt\right)\,.
\end{equation}

\noindent In Eq. \eqref{SP8} $N$ and $N^\alpha{}$ are the lapse\index{lapse} and shift\index{shift} functions respectively, $h_{\alpha\beta}$ is the induced metric\index{metric!induced} on $\Sigma_t{}$ and one can show that
\begin{equation}
\label{SP9}
\sqrt{-g}=N\sqrt{h}.
\end{equation}

\noindent Therefore definitions \eqref{SP4} and \eqref{SP7} agree.\\

\noindent We would now like to prove Proposition \ref{conserv2} below.

\begin{proposition}\label{conserv2} 
\begin{equation}
\label{SP11}
D_\mu{}\left(\bar{\phi}\,\bar{\gamma}^\mu{}\psi\right)=0.
\end{equation}
\end{proposition}

\noindent The proof of this proposition is simple provided one has certain preliminary results available.  So we first derive the required results.\\

\noindent We give the derivation implied in ref. \cite{CL76}, Eq. (21), in order to find the expression for $D_{\mu}\bar{\psi}$, where $\psi$ is a solution of the Dirac equation.  We use the fact that $\bar{\psi}\psi$ is a 0-form field, therefore

\begin{align}
D_\mu{}\left(\bar{\psi}\psi\right)&=\left(D_\mu{}\bar{\psi}\right)\psi+\bar{\psi}D_\mu{}\psi\,,\label{sc1}\\\nonumber\\
&=\left(D_\mu{}\bar{\psi}\right)\psi+\bar{\psi}\left(I\partial_{\mu}+\Gamma_{\mu}\right)\psi\,,\label{sc2}\\\nonumber\\
&\equiv\left(I\partial_{\mu}\bar{\psi}+\mathbb{G}_{\mu}\bar{\psi}\right)\psi+\bar{\psi}\partial_{\mu}\psi+\bar{\psi}\Gamma_{\mu}\psi\,,\label{sc3}\\\nonumber\\
&=\left(\partial_{\mu}\bar{\psi}\right)\psi+\bar{\psi}\partial_{\mu}\psi\label{sc4}\,.
\end{align}

\noindent Eqs. \eqref{sc3} and \eqref{sc4} imply that
\begin{align}
\left(\mathbb{G}_{\mu}\bar{\psi}\right)\psi+\bar{\psi}\Gamma_{\mu}\psi&=0\,,\label{sc5}\\\nonumber\\
\left(\mathbb{G}_{\mu}\bar{\psi}+\bar{\psi}\Gamma_{\mu}\right)\psi&=0\,,\label{sc6}\\\nonumber\\
\mathbb{G}_{\mu}\bar{\psi}+\bar{\psi}\Gamma_{\mu}&=0\,.\label{sc7}
\end{align}

\noindent Thus
\begin{equation}
\label{sc8}
\mathbb{G}_{\mu}\bar{\psi}=-\bar{\psi}\Gamma_{\mu}\,,
\end{equation}

\noindent so, finally, using Eqs. \eqref{sc2}, \eqref{sc3}, and \eqref{sc8} we may write
\begin{equation}
\label{sc9}
D_{\mu}\bar{\psi}=I\partial_{\mu}\bar{\psi}-\bar{\psi}\Gamma_{\mu}\,.
\end{equation}

\noindent Now we would like to find the adjoint of the Dirac equation in curved spacetime, i.e.,
\begin{equation}
\label{sc10}
\left(\bar{\gamma}^\mu{}D_{\mu}\psi+im\psi\right)^{\dagger}=0\,.
\end{equation}

\noindent The derivation requires three steps.  The first step to note that
\begin{equation}
\label{sc11}
\left(\bar{\gamma}^\mu{}\right)^{\dagger}=e_A{}^\mu{}\left(\gamma^A{}\right)^{\dagger}=e_A{}^\mu{}\gamma^0{}\gamma^A{}\gamma^0{}=\gamma^0{}\bar{\gamma}^\mu{}\gamma^0{}\,.
\end{equation}

\noindent The second step is to recall the expression for $\Gamma_{\mu}$ and obtain the result below
\begin{align}
\Gamma_{\mu}^{\dagger}&=\frac{1}{4}\omega_{AB\mu}\left(\gamma^A{}\gamma^B{}\right)^{\dagger}\,,\label{sc12}\\\nonumber\\
&=\frac{1}{4}\omega_{AB\mu}\left(\gamma^B{}\right)^{\dagger}\left(\gamma^A{}\right)^{\dagger}\,,\label{sc13}\\\nonumber\\
&=\frac{1}{4}\omega_{AB\mu}\left(\gamma^0{}\gamma^B{}\gamma^0{}\gamma^0{}\gamma^A{}\gamma^0{}\right)\,,\label{sc14}\\\nonumber\\
&=\frac{1}{4}\omega_{AB\mu}\left(\gamma^0{}\gamma^B{}\gamma^A{}\gamma^0{}\right)\,,\label{sc15}\\\nonumber\\
&=-\frac{1}{4}\omega_{AB\mu}\left(\gamma^0{}\gamma^A{}\gamma^B{}\gamma^0{}\right)\,,\label{sc16}\\\nonumber\\
&=-\gamma^0{}\Gamma_{\mu}\gamma^0{}\,,\label{sc17}
\end{align}

\noindent In the third and final step we make use of Eqs. \eqref{sc11} and \eqref{sc17} and re-write Eq. \eqref{sc10} as follows,
\begin{align}
\left(D_{\mu}\psi\right)^{\dagger}\left(\bar{\gamma}^\mu{}\right)^{\dagger}-im\psi^{\dagger}&=0\,,\label{sc18}\\\nonumber\\
\left[\left(I\partial_{\mu}+\Gamma_{\mu}\right)\psi\right]^{\dagger}\left(\bar{\gamma}^\mu{}\right)^{\dagger}-im\psi^{\dagger}&=0\,,\label{sc19}\\\nonumber\\
\left(\partial_{\mu}\psi^{\dagger}+\psi^{\dagger}\Gamma_{\mu}^{\dagger}\right)\left(\bar{\gamma}^\mu{}\right)^{\dagger}-im\psi^{\dagger}&=0\,,\label{sc20}\\\nonumber\\
\left(\partial_{\mu}\psi^{\dagger}\right)\gamma^0{}\bar{\gamma}^\mu{}\gamma^0{}-\psi^{\dagger}\gamma^0{}\Gamma_{\mu}\gamma^0{}\gamma^0{}\bar{\gamma}^\mu{}\gamma^0{}-im\psi^{\dagger}&=0\,,\label{sc21}\\\nonumber\\
\left(\partial_{\mu}\bar{\psi}\right)\bar{\gamma}^\mu{}\gamma^0{}-\bar{\psi}\Gamma_{\mu}\bar{\gamma}^\mu{}\gamma^0{}-im\psi^{\dagger}&=0\,,\label{sc22}\\\nonumber\\
\left(\partial_{\mu}\bar{\psi}\right)\bar{\gamma}^\mu{}-\bar{\psi}\Gamma_{\mu}\bar{\gamma}^\mu{}-im\bar{\psi}&=0\,,\label{sc23}\\\nonumber\\
\left(I\partial_{\mu}\bar{\psi}-\bar{\psi}\Gamma_{\mu}\right)\bar{\gamma}^\mu{}-im\bar{\psi}&=0\,,\label{sc24}\\\nonumber\\
\left(D_{\mu}\bar{\psi}\right)\bar{\gamma}^\mu-im\bar{\psi}&=0\,.\label{sc25}
\end{align}

\noindent Finally, we return to the proof of Proposition 3.  We assume that $\phi$ and $\psi$ satisfy the Dirac Eq. \eqref{FI13b} and its adjoint Eq. \eqref{sc25}, \index{Dirac equation!adjoint} namely,
\begin{equation}
\label{SP12}
i\bar{\gamma}^\mu{}D_\mu{}\psi-m\psi=0,\;\;\; i\left(D_\mu{}\bar{\psi}\right)\bar{\gamma}^\mu{}+m\bar{\psi}=0,
\end{equation}

\noindent from which it follows that,
\begin{equation}
\label{SP12a}
\bar{\gamma}^\mu{}D_\mu{}\psi=-im\psi,\;\;\; \left(D_\mu{}\bar{\psi}\right)\bar{\gamma}^\mu{}=im\bar{\psi}.
\end{equation}

\noindent The proof of Proposition \ref{conserv2} now follows by applying the Leibniz rule\index{Leibniz rule} to Eq. \eqref{SP11} and using Eqs. \eqref{SP12a} and \eqref{10}.

\begin{proposition}\label{conserv1}\index{Propositions} The scalar product,\index{scalar product!in GR} $\left(\phi\,\vert\,\psi\right)$, Eq. \eqref{SP7}, is conserved, i.e.,
\begin{equation}
\label{SP10}
\frac{d}{dt}\left(\phi\,\vert\,\psi\right)=0\,,
\end{equation}

\noindent where $t=x^0{}$, and provided $\phi$ and $\psi$ satisfy the Dirac Eq. \eqref{FI13} and its adjoint.\index{Dirac equation!adjoint}
\end{proposition}

\noindent Proposition \ref{conserv1} follows from Proposition \ref{conserv2}\index{Propositions} provided, as stated in ref. \cite{P80}, ``we assume that $\phi$ and $\psi$, vanish sufficiently rapidly at spatial infinity or obey suitable boundary conditions in a closed universe,\index{boundary conditions!in closed universe} so that the spatial components of Eq. \eqref{SP11} give vanishing contributions upon integration and the various products are well defined'' (c.f. Eq. \eqref{CJ3}).  For more details refer to \cite{P80} and \cite{HP09}.  

\subsection[Example. The closed FRW universe]{Example. The closed FRW universe}\label{FRW Ex}

\noindent In this section we go over the example from Finster and Reintjes, ref. \cite{FR09}.  We consider the closed FRW universe\index{universe!FRW} whose line element, in conformal coordinates, is (this is in lapse and shift form, Eq. \eqref{SP8})
\begin{equation}
\label{ESP1}
ds^{2}=S(\eta)^{2}\left(d\eta^{2}-d\chi^{2}-f(\chi)^{2}\left(d\theta^{2}+\sin^{2}{\theta}\,d\phi^{2}\right)\right)\,.
\end{equation}

\noindent In the metric \eqref{ESP1} $\eta$ is the conformal time,\index{conformal!time} $\chi$ is the radial coordinate, and $\theta\in (0,\pi)$, $\phi\in [0,2\pi)$, are the angular coordinates.  The scale function $S(\eta)$ depends on the type of matter under consideration.  In the present example $S(\eta)$ is left unspecified and is an arbitrary positive function.  We have
\begin{equation}
\label{ESP1a}
f(\chi)=
\begin{cases}
\sin(\chi), &\text{closed universe,}\;\;\;\chi\in (0,\pi)\\
\sinh{(\chi)}, &\text{open universe,}\;\;\;\;\;\chi>0\,.\\ 
\chi, &\text{flat universe,}\,\;\;\;\;\;\;\;\chi>0\,.\\ 
\end{cases}
\end{equation}

\noindent \noindent  We choose the tetrad vectors\index{tetrad!vectors} in the Cartesian gauge\index{gauge!Cartesian}\index{Cartesian gauge} (see Sec. \ref{Cartesian}).  In this example these tetrad vectors correspond to a class of static observers on the timelike path $\sigma(s)=(\int_{0}^{s}\frac{ds}{S(\eta(s))},\chi_{0},\theta_{0},\phi_{0})$.  One can show that $\nabla_{e_0{}}e_A{}=0$, for all $A$ and any well-behaved $S(\eta)$, so the tetrad $e_A{}$ is a Fermi tetrad field\index{tetrad!Fermi}, Eq. \eqref{FT5}.
\begin{align}
e_{0}&=\frac{1}{S(\eta)}\,\partial_{\eta},\label{ESP2}\\\nonumber\\
e_{1}&=\frac{\sin{\theta}\cos{\phi}}{S(\eta)}\,\partial_\chi{}+\frac{\cos{\theta}\cos{\phi}}{S(\eta)f(\chi)}\,\partial_\theta{}-\frac{\sin{\phi}}{S(\eta)f(\chi)\sin{\theta}}\,\partial_\phi{},\label{ESP3}\\\nonumber\\
e_{2}&=\frac{\sin{\theta}\sin{\phi}}{S(\eta)}\,\partial_\chi{}+\frac{\cos{\theta}\sin{\phi}}{S(\eta)f(\chi)}\,\partial_\theta{}+\frac{\cos{\phi}}{S(\eta)f(\chi)\sin{\theta}}\,\partial_\phi{},\label{ESP4}\\\nonumber\\
e_{3}&=\frac{\cos{\theta}}{S(\eta)}\,\partial_\chi{}-\frac{\sin{\theta}}{S(\eta)f(\chi)}\,\partial_\theta{},\label{ESP5}
\end{align}

\noindent and using Eq. \eqref{FI1b} we obtain the spacetime-dependent gamma matrices,\index{gamma matrices!spacetime dependent}
\begin{align}
\bar{\gamma}^\eta{}&=\frac{1}{S(\eta)}\,\gamma^0{},\label{ESP6}\\\nonumber\\
\bar{\gamma}^\chi{}&=\frac{1}{S(\eta)}\left(\sin{\theta}\cos{\phi}\,\gamma^1{}+\sin{\theta}\sin{\phi}\,\gamma^2{}+\cos{\theta}\,\gamma^3{}\right),\label{ESP7}\\\nonumber\\
\bar{\gamma}^\theta{}&=\frac{1}{S(\eta)f(\chi)}\left(\cos{\theta}\cos{\phi}\,\gamma^1{}+\cos{\theta}\sin{\phi}\,\gamma^2{}-\sin{\theta}\,\gamma^3{}\right),\label{ESP8}\\\nonumber\\
\bar{\gamma}^\phi{}&=\frac{1}{S(\eta)f(\chi)\sin{\theta}}\left(-\sin{\phi}\,\gamma^1{}+\cos{\phi}\,\gamma^2{}\right).\label{ESP9}
\end{align}

\noindent In Eqs. \eqref{ESP6} - \eqref{ESP9} the $\gamma^A{}$ are the constant $\gamma$ matrices in the standard representation\index{representation!standard}.\\

\noindent We write the Dirac equation\index{Dirac equation!in FRW universe} as in \cite{FR09}
\begin{equation}
\label{ESP10}
\left[i\bar{\gamma}^\eta{}\left(\partial_{\eta}+\frac{3}{2}\frac{\dot{S}}{S}\right)+i\bar{\gamma}^\chi{}\left(\partial_{\chi}+\frac{f^{\prime}-1}{f}\right)+i\bar{\gamma}^\theta{}\partial_{\theta}+i\bar{\gamma}^\phi{}\partial_{\phi}-m\right]\Psi=0,
\end{equation}

\noindent where $\dot{S}$ is the derivative with respect to $\eta$, and $f^{\prime}$ is the derivative with respect to $\chi$.  Note that $3\dot{S}/(2S)\neq\Gamma_\eta{}$, etc.  What happens here is similar to what happened in deriving Eq. \eqref{dS4}.\\

\noindent At this point Finster and Reintjes restrict themselves to the closed universe case, $f(\chi)=\sin{(\chi)}$, and assume a solution of the form
\begin{equation}
\label{ESP11}
\Psi(\eta,\chi,\theta,\phi)=\frac{1}{S(\eta)^{\frac{3}{2}}}{h_{1}(\eta)\,\psi_{\lambda}(\chi,\theta,\phi)\choose h_{2}(\eta)\,\tilde{\psi}_{\lambda}(\chi,\theta,\phi)}\,.
\end{equation}

\noindent We shall examine the probability integral\index{probability integral}, Eq. \eqref{SP6}, using the above solution of the Dirac equation in a closed FRW universe\index{Dirac equation!in FRW universe}.  We choose $\Sigma$ to be a slice of constant conformal time\index{conformal!time} $\eta$, then the future-directed normal 1-form $n$ has components (cf. Eq. \eqref{ESP2}),
\begin{equation}
\label{ESP12}
\left(n_\eta{},n_\chi{},n_\theta{},n_\phi{}\right)=\left(S(\eta),0,0,0\right).
\end{equation}

\noindent Using Eqs. \eqref{SP6} and \eqref{ESP6}, we have
\begin{align}
\left(\Psi\,\vert\,\Psi\right)&=\int_{\Sigma}\bar{\Psi}\,\bar{\gamma}^\alpha{}\Psi\,n_\alpha{}\,d\Sigma\,,\label{ESP13}\\\nonumber\\
&=\int_{\Sigma}\bar{\Psi}\,\bar{\gamma}^\eta{}\Psi\,n_\eta{}\,d\Sigma\,,\label{ESP14}\\\nonumber\\
&=\int_{\Sigma}\Psi^{\dagger}\Psi\,d\Sigma\,.\label{ESP15}
\end{align}

\noindent In going from Eq. \eqref{ESP13} to Eq. \eqref{ESP15}, we used Eqs. \eqref{ESP6} and \eqref{ESP12}.  Also 
\begin{equation}
\label{ESP16}
d\Sigma=\sqrt{\vert g_{\Sigma} \vert}\,d\chi\,d\theta\,d\phi=S^{3}(\eta)\,d\mu_{S^{3}}\,.
\end{equation}

\noindent In Eq. \eqref{ESP16}, $g_{\Sigma}$ is the determinant of the induced metric\index{metric!induced} on $\Sigma$, and
\begin{equation}
\label{ESP17}
d\mu_{S^{3}}=\sin^{2}(\chi)\sin(\theta)\,d\chi\,d\theta\,d\phi,
\end{equation}

\noindent is the volume element on the unit sphere $S^{3}$ in hyperspherical coordinates\index{hyperspherical coordinates}.  So substituting Eq. \eqref{ESP11} in Eq. \eqref{ESP13}, we obtain
\begin{align}
\left(\Psi\,\vert\,\Psi\right)&=\int_{\Sigma}\Psi^{\dagger}\Psi\,d\Sigma\,,\label{ESP18}\\\nonumber\\
&=\vert h_{1}\vert ^{2}\int_{S^{3}}\vert\psi_{\lambda}\vert ^{2} d\mu_{S^{3}}+\vert h_{2}\vert ^{2}\int_{S^{3}}\vert\tilde{\psi}_{\lambda}\vert ^{2} d\mu_{S^{3}}\,,\label{ESP19}\\\nonumber\\
&=\vert h_{1}\vert ^{2}+\vert h_{2}\vert ^{2}\,,\label{ESP20}
\end{align}

\noindent where we have followed the normalization of ref. \cite{FR09} (cf. Eq. \eqref{N19}).  It follows from Eq. \eqref{SP10} that the probability integral\index{probability integral} is constant in time.

\begin{appendices}

\section[Tetrads]{Tetrads}\label{Tet}

\subsection[The tetrad formalism]{The tetrad formalism}\label{GenT}

\noindent A tetrad\index{tetrad!vectors} is a set of four linearly independent vectors that can be defined at each point in a (semi -) Riemannian\index{Riemannian} spacetime.  Here we give a summary of useful relations for tetrad\index{tetrad!field} \textit{fields}.  Good detailed discussions can be found in several texts see, for example, appendix J of \cite{C04}.  We have the following basic relations that determine the vector fields $e_A{}^{\alpha}$ or the 1-forms (covector fields) $e^A{}_{\alpha}$, (we may use the notation, $e^{A}=e^A{}_{\alpha}\,dx^{\alpha}$ and $e_{A}=e_A{}^{\alpha}\,\partial_{\alpha}$).  The tetrads by definition satisfy the relations (see \cite{C04}, Eq. (J.3))
\begin{align}
e^A{}_{\alpha}\,e_A{}^{\beta}&=\delta_\alpha{}^{\beta}\,,\label{T1a}\\\nonumber\\
e^A{}_{\alpha}\,e_B{}^{\alpha}&=\delta^A{}_{B}\,.\label{T1b}
\end{align}

\noindent The choice of the tetrad field\index{tetrad!field} determines the metric through Eq. \eqref{T1} below.
\begin{align}
g_{\alpha\beta}&=e^A{}_{\alpha}\,e^B{}_{\beta}\,\eta_{AB}\,,\label{T1}\\\nonumber\\
\eta_{AB}&=e_A{}^{\alpha}\,e_B{}^{\beta}\,g_{\alpha\beta}\,,\label{T2}
\end{align}

\noindent where $\eta_{AB}$ is the Minkowski\index{Minkowski} spacetime metric in Cartesian coordinates. We shall always assume that the velocity vector field\index{velocity}, $e_0{}$, is tangent to a congruence of timelike paths\index{congruence!of timelike paths} and thus the tetrads are moving along these paths.  The reader should also read the comments in Sec. \ref{FI I} below Eq. \eqref{FI1b}, and in Sec. \ref{NP} below Eq. \eqref{NP10.4}.\\

\noindent Under coordinate transformations, greek indices are treated as tensor indices, while latin indices are merely labels (thus the $e_A{}^{\alpha}$ represent four different vector fields).  Equations \eqref{T2} are also a statement of the orthonormality of the vectors $e_A{}^{\alpha}$. The tetrad components may be determined using the Eqs. \eqref{T1} or \eqref{T2}.

\begin{remark}\label{Label}\index{Remarks}  It is easy to convince oneself that relabeling the subscripts \textup{(}or superscripts\textup{)} of the $e_A{}^\alpha{}$ in a consistent way, does not affect the relation \eqref{T2}.  However, problems may arise, if one is careless with relabeling and reordering variables while using a symbolic manipulation software.
\end{remark}

\noindent Although in these notes we have considered spacetimes with dimensionality of two or four, in general, if $n$ is the dimensionality of the manifold, Eqs. \eqref{T2} are a set of $(\frac{1}{2})n(n+1)$ equations for the $n^{2}$ unknown components of the \textit{vielbein} $e_A{}^{\alpha}$.  Therefore $(\frac{1}{2})n(n-1)$ components can be freely chosen or determined by extra conditions.

\begin{exercise}\label{TFe1}\index{Exercises} It is a simple exercise to show that Eqs. \eqref{T1b} and \eqref{T1}, imply Eq. \eqref{T2}, while Eqs. \eqref{T1a} and \eqref{T2}, imply Eq. \eqref{T1}.
\end{exercise}

\noindent  We have the following rules for raising and lowering indices,
\begin{align}
e_{A\alpha}&=g_{\alpha\beta}\,e_A{}^{\beta}\,,\label{T4}\\\nonumber\\
e^A{}_{\alpha}&=\eta^{AB}e_{B\alpha}\,.\label{T5}
\end{align}

\noindent The components of tensors in the tetrad frame\index{tetrad!frame} are given by relations such as the ones
below 
\begin{equation}
\label{T6}
V^{A}=e^A{}_{\alpha}V^{\alpha},
\end{equation}
\begin{equation}
\label{T7}
T^A{}_{B}=e^A{}_{\alpha}\,e_B{}^{\beta}\,T^\alpha{}_{\beta}\,,
\end{equation}

\noindent and so on.  Note that in Eq. \eqref{T6} we are taking the product of a vector $V$ with $n$ 1-forms $e^{A}_{\;\;\;\alpha}$, as a result, we are replacing the vector $V$ with $n$ scalars $V^{A}$.  Likewise in Eq. \eqref{T7}, we are replacing the $\binom{1}{1}$ tensor T with $n^{2}$ scalars $T^{A}_{\;\;\;B}$ \cite{W72}.\\

\noindent We can obtain the tensor components in the global chart from the ``components'' in the tetrad frame\index{tetrad!frame} using relations like the one below
\begin{equation}\label{T8}
V^{\alpha}=e_A{}^{\alpha}\,V^{A}=e^{A\alpha}\,V_{A}\,.
\end{equation}

\noindent Using the above relations we can show that $U_{\mu}V^{\mu}=U_{A}V^{A}$.
\begin{align}
g_{\mu\nu}U^{\mu}V^{\nu}&=\eta_{AB}e^A{}_{\mu}e^B{}_{\nu}\,U^{\mu}V^{\nu},\label{T9}\\
g_{\mu\nu}U^{\mu}V^{\nu}&=\eta_{AB}U^{A}V^{B},\label{T10}\\
U_{\mu}V^{\mu}&=U_{A}V^{A}\label{T11}.
\end{align}

\subsection[Fermi tetrad fields]{Fermi tetrad fields}\label{Fermi}

\noindent  A Fermi\index{tetrad!Fermi} tetrad field\index{tetrad!field} must satisfy some special conditions.  As usual the tetrad field\index{tetrad!field}, satisfies Eq. \eqref{T1}, etc., but in the Fermi case, the velocity vector field, $e_0{}$, is tangent to a congruence of timelike geodesics,\index{congruence! of timelike geodesics} $\sigma(\tau)$, parametrized by the proper time $\tau$.  Thus
\begin{equation}
\label{FT1}
e_0{}=\frac{d\sigma(\tau)}{d\tau}\,,
\end{equation}

\noindent and therefore,
\begin{equation}
\nabla_{e_{0}}e_0{}=0\,.\label{FT4}
\end{equation}

\noindent A Fermi tetrad\index{tetrad!Fermi} field must satisfy the equations
\begin{equation}
\nabla_{e_{0}}e_A{}=0,\label{FT5}\;\;\;\;A=0,1,2,3,
\end{equation}

\noindent so that all the tetrad vectors are parallelly transported along the chosen congruence of timelike geodesics\index{congruence!of timelike geodesics}.\\

\noindent \textup{(}Recall that if $\nabla_u{}u\neq 0$ but $\nabla_u{}v=0$, then $v$ is parallel transported but not on a geodesic\textup{)}.

\begin{remark}\label{FermiTet}\index{Remarks}  Given a Fermi tetrad\index{tetrad!Fermi} enables one to obtain approximate Fermi coordinates\index{Fermi coordinates} by the well-known process given in \textup{\cite{KC08}}.  Examples of how to obtain exact Fermi coordinates, in cases where this is possible, were given in \textup{\cite{CM06}, \cite{KC10}, \cite{BGJ11}, \cite{K12}}.\end{remark}

\subsection[The Cartesian gauge tetrad]{The Cartesian gauge tetrad}\label{Cartesian}\index{Cartesian gauge}

\noindent A tetrad field referred to as the ``Cartesian gauge''\index{tetrad!Cartesian gauge}\index{Cartesian gauge} was introduced by Brill and Wheeler \cite{BW57}, and has been found useful by many authors \cite{VP90} - \cite{CCS16}).  An example of the Cartesian gauge tetrad\index{tetrad!Cartesian gauge}\index{Cartesian gauge} used in the FRW universe\index{universe!FRW} was given above, Eqs. \eqref{ESP2} - \eqref{ESP5}.  Here we first discuss it in the simplest case namely flat Lorentz spacetime\index{Lorentz!spacetime}.  Consider the standard tetrad 1-forms\index{tetrad!1-forms} in (Cartesian) Minkowski spacetime\index{Minkowski}, namely,
\begin{align}
e^{0}&=dt\,,&e^{1}=dx\,,\nonumber\\\label{CT1}\\
e^{2}&=dy\,,&e^{3}=dz\,.\nonumber
\end{align}

\noindent Then, if we transform to spherical coordinates $(t,r,\theta,\phi)$, the above 1-forms transform into the 1-form tetrad below
\begin{align}
\omega^0{}&=dt\,,\label{CT2}\\\nonumber\\
\omega^1{}&=\sin{\theta}\cos{\phi}\,dr+r\cos{\theta}\cos{\phi}\,d\theta-r\sin{\theta}\sin{\phi}\,d\phi\,,\label{CT3}\\\nonumber\\
\omega^2{}&=\sin{\theta}\sin{\phi}\,dr+r\cos{\theta}\sin{\phi}\,d\theta+r\sin{\theta}\cos{\phi}\,d\phi,\label{CT4}\\\nonumber\\
\omega^3{}&=\cos{\theta}\,dr-r\sin{\theta}\,d\theta\,,\label{CT5}
\end{align}

\noindent For the case of the flat spacetime the Fock-Ivanenko coefficients\index{Fock-Ivanenko coefficients} vanish with this tetrad in spherical coordinates.  The metric is the usual spherical coordinate metric,
\begin{equation}
\label{CT6}
ds^{2}=dt^{2}-dr^{2}-r^{2}(d\theta^{2}+\sin^{2}\theta\,d\phi^{2})\,.
\end{equation}

\noindent Note that the default 1-form tetrad for the metric of Eq. \eqref{CT6}, namely,
\begin{align}
\omega^0{}&=dt\,,\label{CT7}\\\nonumber\\
\omega^1{}&=dr\,,\label{CT8}\\\nonumber\\
\omega^2{}&=r\,d\theta,\label{CT9}\\\nonumber\\
\omega^3{}&=r\sin{\theta}\,d\phi\,,\label{CT10}
\end{align}

\noindent is ``rotated'' (see Appendix \ref{V&LB}) with respect to the original Cartesian tetrad Eq. \eqref{CT1} and the Fock-Ivanenko coefficients no longer vanish.\\

\begin{exercise}\label{CGe1}\index{Exercises} \textup{(a)} Write the Dirac equation using the tetrad, \eqref{CT2} - \eqref{CT5}, for the metric \eqref{CT6}, in the chiral representation\index{representation!chiral} Eq. \eqref{D6}.  \textup{(b)} Transform the chiral plane wave solutions Eq. \eqref{C9} to spherical coordinates and show that they satisfy the Dirac equation obtained in part \textup{(a)}.
\end{exercise}

\subsection[Vielbeins, spinors and Lorentz matrices]{Vielbeins, spinors and Lorentz matrices}\label{V&LB}

\noindent In this section we present the results without proofs and we refer the reader to the proofs given in \cite{L76}.  We mention for the sake of clarity that if $F \in G$, where $G$ is a group of coordinate transformations, then we write
\begin{equation}
\label{1S1}
\bar{x}=F x\,.
\end{equation}

\noindent Thus in general for a scalar function, $\phi(x)$, we have
\begin{equation}
\label{1S2}
\phi(x)=\bar{\phi}(\bar{x}).
\end{equation}

\noindent If in a coordinate system \textcolor{red}{$(x^0{},x^{i})$,} we change from an initial chosen vielbein\footnote {We shall use the term vielbein\index{vielbein} whenever the dimensionality is not necessarily (3+1).  We reserve the term tetrad for the (3+1) case.} set, $h_A{}$, to another set, $e_A{}$, then the new vielbein vectors can be expressed as linear combinations of the old\footnote {A shorthand for Eq. \eqref{1S3} is $e=\Lambda^{-1}h$, while Eq. \eqref{1S7} is $L^{-1}\gamma L=\Lambda\gamma$, so that our index positions agree with refs. \cite{W84}, \cite{R96} and \cite{N92}.},
\begin{equation}
\label{1S3}
e_A{}^\mu{}=\Lambda_A{}^B{}h_B{}^\mu{}\,.
\end{equation}

\noindent However, both vielbein sets must satisfy Eq. \eqref{T2}, i.e.,
\begin{align}
\eta_{AB}&=h_A{}^{\alpha}\,h_B{}^{\beta}\,g_{\alpha\beta}\,,\label{1S3a}\\\nonumber\\
\eta_{AD}&=e_A{}^{\mu}\,e_D{}^{\nu}\,g_{\mu\nu}\,.\label{1S3b}
\end{align}

\noindent Substituting Eq. \eqref{1S3} in Eq. \eqref{1S3a}, we obtain
\begin{align}
\Lambda_A{}^B{}\,h_B{}^\mu{}\,\Lambda_D{}^C{}\,h_C{}^\nu{}\,g_{\mu\nu}&=\eta_{AD}\,,\label{1S3c}\\\nonumber\\
\Lambda_A{}^B{}\Lambda_D{}^C{}\eta_{BC}&=\eta_{AD}\,,\label{1S3d}\\\nonumber\\
\Lambda^{T}\eta\,\Lambda&=\eta\,,\label{1S3e}
\end{align}

\noindent where $\Lambda^{T}$ is the transpose of $\Lambda$.  From Eq. \eqref{1S3e} we have that $\det{\Lambda}=\pm 1$.  Thus it follows then from Eq. \eqref{1S3d} that $\Lambda_A{}^B{}$ is a Lorentz matrix.\index{Lorentz!matrix}  So in the context of general relativity \textit{the Lorentz group\index{Lorentz!group} is the group of vielbein\index{vielbein} rotations} \cite{L76} p. 143.  We also remark that the $\Lambda$ matrices will in general be spacetime-dependent.\\

\noindent Under coordinate transformations spinors, $\psi$, behave like scalars so that, \cite{L76} p. 147,
\begin{equation}
\label{1S4}
\left(\begin{array}{c}\bar{\phi}(\bar{x})\\
\bar{\chi}(\bar{x})
\end{array}\right)=\left(\begin{array}{c}\phi(x)\\
\chi(x)
\end{array}\right).
\end{equation}

\noindent However when a vielbein\index{vielbein} $h_B{}$, is rotated by $\Lambda$ as in Eq. \eqref{1S3}, then
\begin{equation}
\label{1S5}
\psi_{e}=L\,\psi_{h}\,,
\end{equation}

\noindent where $L$ is a (spacetime-dependent) spinor representative of a vielbein\index{vielbein} rotation $\Lambda$, \cite{L76} pp. 76, 147,
\begin{equation}
\label{1S6}
L=\left(
 \begin{matrix}
     S & 0\\
     \rule{0in}{2ex}
     0 & \left(S^\dagger{}\right)^{-1}\\
 \end{matrix}
  \right),
\end{equation}

\noindent with $\det(L)=1$, \cite{W84}, that satisfies the relations, \cite{L76} p. 147,
\begin{equation}
\label{1S7}
L^{-1}\gamma^A{}L=\Lambda^A{}_B{}\gamma^B{}\,,
\end{equation}

\noindent and
\begin{equation}
\label{1S8}
\gamma^0{}L^{\dagger}\gamma^0{}=L^{-1}\,,
\end{equation}

\noindent Given in \cite{PT09}, Eq. (5.396), p.246.  For a derivation of Eq. \eqref{1S7} see, e.g., \cite{R96}.

\section[The gamma matrices]{The gamma matrices}\label{Gam}

\subsection[General summary]{General summary}\label{Gen}

\noindent The $2\times 2$ Pauli spin matrices\index{Pauli spin matrices} are
\begin{equation}
\label{D0}
\sigma^{1}=\left(
 \begin{matrix}
     0 &  1\\
     \rule{0in}{2ex}
     1 & 0\\
 \end{matrix}
 \right),\;\;\;\;\sigma^{2}=\left(
 \begin{matrix}
     0 &  -i\\
     \rule{0in}{2ex}
     i & 0\\
 \end{matrix}
 \right),\;\;\;\;\;
 \sigma^{3}=\left(
 \begin{matrix}
     1 &  0\\
     \rule{0in}{2ex}
    0 & -1\\
 \end{matrix}
 \right).
\end{equation}

\noindent For a free spin $1/2$ particle of mass $m$ we write the Dirac equation in Minkowski spacetime\index{Dirac equation!in Minkowski spacetime} as
\begin{equation}
\label{D1}
i\gamma^{A}\partial_{A}\psi-m\psi=0\,,
\end{equation}

\noindent where $\psi$ is a 4-component (contravariant) spinor\index{spinor!4-component} and the $4\times 4$ $\gamma$ (constant) matrices \footnote {So more precisely Eq. \eqref{D1} is $i(\gamma^A{})^i{}_{k}\,\partial_A{}\psi^{k}-m\psi^{i}=0,\;i,k = (1,2,3,4)$.}, satisfy the anticommutation relation\index{anticommutation relation} 
\begin{equation}
\label{D2}
\{\gamma^{A},\gamma^{B}\}=\varepsilon\, 2\eta^{AB}I\,,
\end{equation}

\noindent where $\varepsilon=\pm 1$, and the Hermiticity conditions\index{Hermiticity conditions}
\begin{equation}
\label{D3}
(\gamma^{A})^{\dagger}=\gamma^{0}\gamma^{A}\gamma^{0}\,.
\end{equation}

\noindent We raise and lower the indices using the metric $\eta$, e.g., $\gamma^{A}=\eta^{AB}\gamma_{B}$.\\

\noindent With the exception of the Jauch-Rohrlich representation\index{representation!Jauch-Rohrlich} below, the other three representations are given in a form so that they satisfy the relation\begin{equation}
\label{D10}
\{\gamma^{A},\gamma^{B}\}=2\eta^{AB}I\,,
\end{equation}

\noindent with signature convention $(+, -, -, -)$.  We also define the matrix
\begin{equation}
\label{D8}
\gamma^{5}:=i\,\gamma^{0}\gamma^{1}\gamma^{2}\gamma^{3}\,,
\end{equation}

\noindent which satisfies the representation and signature independent relations,
\begin{equation}
\label{D9}
\{\gamma^{A},\gamma^{5}\}=0\,,\;\;\;\;\left(\gamma^{5}\right)^{2}=I_{4}\,,\;\;\;\;\left(\gamma^{5}\right)^{\dagger}=\gamma^{5}\,.
\end{equation}

\begin{remark}\label{sign 1}\index{Remarks} The sign choice in \textup{Eq.} \eqref{D2} depends on the metric sign convention and the representation of the $\gamma$ matrices.  There are several commonly used representations each with its own advantages.  One can avoid the $(\varepsilon=-1)$ choice in \textup{Eq.} \eqref{D2} by multiplying the $\gamma$ matrices with $\pm i$, \textup{(}e.g., both \textup{\cite{P80}} and \textup{\cite{CRR12}} multiply by $-i$\textup{)}.\end{remark}

\begin{remark}\label{sign 2}\index{Remarks} We also point out that the $(-)$ sign in front of the mass $m$ in the Dirac equation \eqref{D1}\index{Dirac equation!mass sign}, can be changed to a $(+)$ by multiplying the Dirac equation \textup{(}from the left\textup{)} by $\gamma^{5}$.  One finds that the spinor $\gamma^{5}\psi$ obeys the Dirac equation with the opposite sign in the mass term.  This is true also in curved spacetime, see \eqref{FI13}. \end{remark}

\subsection[The standard representation]{The standard or Dirac-Pauli representation}\label{D-P}

In the \textit{standard or Dirac-Pauli representation\index{representation!Dirac-Pauli}\index{representation!standard}}, (or the \textit{Bjorken-Drell representation}\index{representation!Bjorken-Drell}) we have
\begin{equation}
\label{D4}
\gamma^{0}=\left(
 \begin{matrix}
     I_{2} &  0\\
     \rule{0in}{2ex}
     0 & -I_{2}\\
 \end{matrix}
 \right),\;\;\;\;\gamma^{K}=\left(
 \begin{matrix}
     0 &  \sigma^{K}\\
     \rule{0in}{2ex}
     -\sigma^{K}& 0\\
 \end{matrix}
 \right),\;\;\;\; K=(1,2,3)\,.
\end{equation}

\noindent It is easy to verify that,
\begin{equation}
\label{D5}
\left(\gamma^{0}\right)^{2}=I\,,\hspace{0.5cm}\left(\gamma^{K}\right)^{2}=-I\,.
\end{equation}

\noindent We also give below the Dirac $\beta$ and $\alpha^{K}$ matrices,
\begin{equation}
\label{D5a}
\beta=\left(
 \begin{matrix}
     I_{2} &  0\\
     \rule{0in}{2ex}
     0 & -I_{2}\\
 \end{matrix}
 \right),\;\;\;\;\alpha^{K}=\left(
 \begin{matrix}
     0 &  \sigma^{K}\\
     \rule{0in}{2ex}
     \sigma^{K}& 0\\
 \end{matrix}
 \right),\;\;\;\; K=(1,2,3)\,,
\end{equation}

\noindent that is, $\gamma^0{}=\beta,\;\gamma^K{}=\beta\alpha^{K}$.

\subsection[The chiral representation]{The chiral or Weyl representation}\label{C}

In the \textit{chiral or Weyl representation}\index{representation!Weyl}\index{representation!chiral}, there are two possible choices for $\gamma^0{}$, we choose 
\begin{equation}
\label{D6}
\gamma^{0}=\left(
 \begin{matrix}
     0 &  -I_{2}\\
     \rule{0in}{2ex}
     -I_{2} & 0\\
 \end{matrix}
 \right),\;\;\;\;\gamma^{K}=\left(
 \begin{matrix}
     0 &  \sigma^{K}\\
     \rule{0in}{2ex}
     -\sigma^{K}& 0\\
 \end{matrix}
 \right),\;\;\;\; K=(1,2,3)\,,
\end{equation}

\noindent and
\begin{equation}
\label{D7}
\left(\gamma^{0}\right)^{2}=I\,,\hspace{0.5cm}\left(\gamma^{K}\right)^{2}=-I\,.
\end{equation}

\noindent  Another option, in the chiral representation\index{representation!chiral}, is to choose the negative of the above $\gamma^0{}$, in which case $\gamma^5{}$ changes sign, unless one defines it as the negative of Eq. \eqref{D8}.  Some authors define the chiral $\gamma$ matrices by multiplying all of the $\gamma$'s in Eq. \eqref{D6} by $(-1)$, then the $\gamma^5{}$ does not change sign.  In any of the above-mentioned chiral representations the $\gamma^{5}$ given by Eq. \eqref{D8} is equal to
\begin{equation}
\label{D8}
\gamma^{5}=\pm\left(
 \begin{matrix}
     I_{2} &  0\\
     \rule{0in}{2ex}
    0& -I_{2}\\
 \end{matrix}
 \right),
\end{equation}

\noindent (see also Sec. \ref{CR}).

\subsection[The Majorana representation]{The Majorana representation}\label{M}

In the \textit{Majorana representation}\index{representation!Majorana} the $\gamma$ matrices are imaginary and the spinors are real.
\begin{align}
\label{M1m}
\gamma^{0}&=\left(
 \begin{matrix}
     0 &  \sigma^{2}\\
     \rule{0in}{2ex}
     \sigma^{2} & 0\\
 \end{matrix}
 \right),\;\;\;\;\gamma^{1}=\left(
 \begin{matrix}
     i\sigma^{3} &  0\\
     \rule{0in}{2ex}
     0& i\sigma^{3}\\
 \end{matrix}
 \right),\\\nonumber\\
\label{M2m}
\gamma^{2}&=\left(
 \begin{matrix}
     0 &  -\sigma^{2}\\
     \rule{0in}{2ex}
     \sigma^{2} & 0\\
 \end{matrix}
 \right),\;\;\;\;\gamma^{3}=\left(
 \begin{matrix}
     -i\sigma^{1} &  0\\
     \rule{0in}{2ex}
     0& -i\sigma^{1}\\
 \end{matrix}
 \right),
\end{align}

\noindent and
\begin{equation}
\label{M3m}
\left(\gamma^{0}\right)^{2}=I\,,\hspace{0.5cm}\left(\gamma^{K}\right)^{2}=-I\,.
\end{equation}

\subsection[The Jauch-Rohrlich representation]{The Jauch-Rohrlich representation}\label{J-R}

In the \textit{Jauch-Rorhlich representation}\index{representation!Jauch-Rohrlich} \cite{JR76}, we have
\begin{equation}
\label{J1}
\gamma^{0}=-i\left(
 \begin{matrix}
     I_{2} &  0\\
     \rule{0in}{2ex}
     0 & -I_{2}\\
 \end{matrix}
 \right),\;\;\;\;\gamma^{K}=\left(
 \begin{matrix}
     0 &  \sigma^{K}\\
     \rule{0in}{2ex}
     \sigma^{K}& 0\\
 \end{matrix}
 \right),\;\;\;\; K=(1,2,3)\,,
\end{equation}

\noindent in fact form Eq. \eqref{D5a} we have that,
\begin{equation}
\label{J1a}
\gamma^0{}=-i\beta\,,\hspace{0.5cm}\gamma^K{}=\alpha^K{}\,,
\end{equation}

\noindent thus,

\begin{equation}
\label{J2}
\left(\gamma^{0}\right)^{2}=-I\,,\hspace{0.5cm}\left(\gamma^{K}\right)^{2}=I\,,
\end{equation}

\noindent and we satisfy
\begin{equation}
\label{J3}
\{\gamma^{A},\gamma^{B}\}=2\eta^{AB}I\,,
\end{equation}

\noindent with signature $(-,+,+,+)$.\\

\noindent Jauch and Rorhlich define $\gamma^{5}\equiv\gamma_5{}\equiv\gamma^{0}\gamma^{1}\gamma^{2}\gamma^{3}$, so again $\gamma^5{}$ satisfies Eqs. \eqref{D9}. 
Furthermore, instead of Eq. \eqref{D1}, we now have
\begin{equation}
\label{J4}
\gamma^{A}\partial_{A}\psi+m\psi=0\,.
\end{equation}

\section[Metric signatures, the FI coefficients, etc.]{Metric signatures, the FI coefficients, etc.}\label{SigCartan}

\noindent For easy reference we begin by recalling our definitions of the spin connection coefficients\index{spin connection}, $\omega_{AB\mu}$, the spinor affine connection,\index{spinor!affine connection} $\Gamma_{\mu}$, the Fock-Ivanenko coefficients,\index{Fock-Ivanenko coefficients} $\Gamma_{C}$, and the anticommutation relations of the $\gamma$ matrices.
\begin{align}
&\omega_{AB\mu}=g_{\beta\alpha}e_A{}^\alpha{}\,\nabla_{\mu}\,e_B{}^{\beta}\,,\label{Sig1}\\\nonumber\\
&\Gamma_{\mu}=\frac{\varepsilon}{4}\,\omega_{AB\mu}\,\gamma^{A}\gamma^{B}\,,\label{Sig2}\\\nonumber\\
&\Gamma_{C}=e_C{}^{\mu}\,\Gamma_{\mu}\,,\label{Sig3}\\\nonumber\\
&\{\gamma^{A},\gamma^{B}\}=\varepsilon\,2\eta^{AB}I\,.\label{Sig4}
\end{align}

\subsection[Signature (-2)]{Signature (-2)}\label{Sig(-2)}

\noindent It is clear that the sign of the $\omega_{AB\mu}$ coefficients depends on the signature\index{signature} because of the $g_{\beta\alpha}$ factor in Eq. \eqref{Sig1}.  We now let $\varepsilon=+1$ in Eqs. \eqref{Sig2}, \eqref{Sig4} and use any $\gamma$ matrix representation whose matrices satisfy
\begin{equation}
\label{Sig5}
\left(\gamma^{0}\right)^{2}=I\,,\hspace{0.5cm}\left(\gamma^{K}\right)^{2}=-I\,,
\end{equation}

\noindent and consequently Eq. \eqref{Sig4}.  We then obtain $\Gamma_\mu{}$ and $\Gamma_C{}$ and we may write the Dirac equation as
\begin{equation}
\label{Sig6}
i\gamma^{C}\left(e_C{}+\Gamma_C{}\right)\,\psi-m\psi=0\,.
\end{equation}

\begin{remark}\label{Cartan FI1}\index{Remarks} Following the above assumptions and steps in the software package \textsc{Cartan}\index{Cartan@\textsc{Cartan}}, one will find that the resulting $\Gamma_C{}$ coefficients have the opposite sign from ours.  This is because the $\Gamma_C{}$ coefficients are defined with the opposite sign in the software.  However, this is compensated in \textsc{Cartan}\index{Cartan@\textsc{Cartan}} by inserting another minus sign so that the Dirac equation is now
\begin{equation}
\label{Sig7}
i\gamma^{C}\left(e_C{}-\Gamma_C{}\right)\,\psi-m\psi=0\,,\;\;\;\mbox{\textup{(}\textsc{Cartan}\textup{)}},
\end{equation}

\noindent thus identical to Eq. \eqref{Sig6}.\end{remark}

\subsection[Signature (+2)]{Signature (+2)}\label{Sig(+2)}

\noindent We begin by again letting $\varepsilon=+1$ in Eqs. \eqref{Sig2}, \eqref{Sig4}.  In order to satisfy Eq. \eqref{Sig4}, we multiply the $\gamma$ matrices by $(+i)$ so that now, instead of Eqs. \eqref{Sig5}, we have
\begin{equation}
\label{Sig8}
\left(\gamma^{0}\right)^{2}=-I\,,\hspace{0.5cm}\left(\gamma^{K}\right)^{2}=I\,.
\end{equation}

\noindent It is easy to see from Eq. \eqref{Sig2} that the change of signature will also change the sign of the coefficients $\omega_{AB\mu}$.  Thus the latter sign change along with the product of the two $(+i)$ factors from the $\gamma$ matrices in Eq. \eqref{Sig2}, will give us the same $\Gamma_C{}$ coefficients as before (Sec. \ref{Sig(-2)}).  The Dirac equation is now written as
\begin{equation}
\label{Sig9}
\gamma^{C}\left(e_C{}+\Gamma_C{}\right)\,\psi-m\psi=0\,,
\end{equation}

\noindent since the $(+i)$ factor has been absorbed in the $\gamma^C{}$.

\begin{remark}\label{Cartan FI2}\index{Remarks} The software package \textsc{Cartan}\index{Cartan@\textsc{Cartan}}, also uses $(+i)$ for this signature, but recall that the $\Gamma_C{}$ coefficients have the opposite sign from ours.  So that \textsc{Cartan}'s\index{Cartan@\textsc{Cartan}} Dirac equation is now
\begin{equation}
\label{Sig10}
\gamma^{C}\left(e_C{}-\Gamma_C{}\right)\,\psi-m\psi=0\,,\;\;\;\mbox{\textup{(}\textsc{Cartan}\textup{)}}.
\end{equation}
\end{remark}

\noindent A number of authors prefer to multiply their $\gamma$ matrices with a factor $(-i)$, e.g., \cite{P80}, \cite{CRR12}.  With the definitions in our paper or the ones in ref. \cite{CRR12} the Dirac equation would be
\begin{equation}
\label{Sig11}
\gamma^{C}\left(e_C{}+\Gamma_C{}\right)\,\psi+m\psi=0\,.
\end{equation}

\noindent Parker in \cite{P80}, using the Dirac $\beta$ and $\alpha^{K}$ matrices, Eq. \eqref{D5a}, has $\gamma^0{}=\eta^{00}\gamma_0{}=-i\beta$, and $\gamma^K{}=\gamma^0{}\alpha^K{}$.  In addition Parker defines his $\Gamma_\mu$ with the opposite sign from the one adopted here and compensates with the usual $(-)$ sign change in the Dirac equation.  Thus his Dirac equation is
\begin{equation}
\label{Sig12}
\gamma^{C}\left(e_C{}-\Gamma_C{}\right)\,\psi+m\psi=0\,.
\end{equation}

\noindent As another example we consider Ryder in ref. \cite{R09}.  Ryder uses $\varepsilon=-1$ in Eqs. in Eqs. \eqref{Sig2}, \eqref{Sig4}, so he can use the usual $\gamma$ matrix representations with
\begin{equation}
\label{Sig13}
\left(\gamma^{0}\right)^{2}=I\,,\hspace{0.5cm}\left(\gamma^{K}\right)^{2}=-I\,.
\end{equation}

\noindent Note, however, that the $\varepsilon=-1$ along with the sign change due to the signature, ultimately gives the same $\Gamma_\mu{}$ and $\Gamma_C{}$ coefficients as ours obtained in Sec. \ref{Sig(-2)}.  Clearly, using $\varepsilon=-1$ is just completely equivalent to multiplying the $\gamma$ matrices with $(\pm i)$, except that now we don't have to hide the $(i)$ in the Dirac equation, which retains its standard form (see \cite{R09}, Eq. (11.129)).
\begin{equation}
\label{Sig14}
i\gamma^{C}\left(e_C{}+\Gamma_C{}\right)\,\psi-m\psi=0\,.
\end{equation}

\noindent Finally one may use the Jauch-Rohrlich representation\index{representation!Jauch-Rohrlich}  (see Secs. \ref{HM} and \ref{C}).

\section[Dirac plane wave solutions in SR]{Dirac plane wave solutions in SR}\label{Dir}

\subsection[Notation]{Notation}\label{Not}

\noindent In the calculations below we use the metric sign convention $(+,-,-,-)$.  As usual $c=\hbar=1$.  Thus we write the Minkowski\index{Minkowski} metric as
\begin{equation}
\label{Not1}
ds^{2}=\eta_{\mu\nu}dx^{\mu}dx^{\nu}=dt^{2}-dx^{2}-dy^{2}-dz^{2}\,,
\end{equation}

\noindent where $\mu$ and $\nu$ run over $(0,1,2,3)$ or $(t,x,y,z)$, Thus there is no distinction between the upper case latin indices used in Sec. \ref{SCD} and the greek indices in the present section \ref{Dir}.  We write $p=\left(p^{0},\boldsymbol{p}\right)$, where,
\begin{align}
p^{0}&=p_{0}\,,\label{Not2}\\
p^{j}&=-p_{j}\,,\;\;\;\;j=(1,2,3)=(x,y,z),\label{Not3}
\end{align}

\noindent and
\begin{align}
p^{0}&=p_{0}=i\partial_{t}\,,\label{Not4}\\
p^{j}&=-p_{j}=-i\partial_{j}\,.\label{Not5}
\end{align}

\subsection[The Dirac equation]{The Dirac equation}\label{DE}

\noindent The Dirac equation for a free, spin $1/2$, particle of mass $m$ in Minkowski\index{Minkowski} spacetime is usually written as 
\begin{equation}
\label{DE1}
i\gamma^{\mu}\partial_{\mu}\psi-m\psi=0\,,
\end{equation}

\noindent and introducing the Feynman ``slash'' notation,\index{Feynman!slash notation}
\begin{equation}
\label{DE2}
\slashed{p}=\gamma^{\mu}p_{\mu}\,,
\end{equation}

\noindent we may rewrite the Dirac Eq. \eqref{DE1} in the shorthand version
\begin{equation}
\label{DE3}
(\slashed{p}-mI)\psi=0,
\end{equation}

\noindent or,
\begin{equation}
\label{DE4}
i\gamma^{0}\partial_{t}\psi+i\gamma^{1}\partial_{x}\psi+i\gamma^{2}\partial_{y}\psi+i\gamma^{3}\partial_{z}\psi-mI\psi=0\,,
\end{equation}

\noindent where $\psi$ is a 4-component spinor,
\begin{equation}
\label{DE5}
\psi=\left(\begin{array}{c}\psi_1{}\\
\psi_2{}\\
\psi_3{}\\
\psi_4{}
\end{array}\right).
\end{equation}

\noindent The $\gamma$ matrices are reviewed in Appendix \ref{Gam}.

\subsection[Plane wave solutions in the standard representation]{Plane wave solutions in the standard representation}\label{PW}

\noindent In this section we will write the plane wave solutions of the Dirac equation, first using the \textit{standard representation}\index{representation!standard} of the $\gamma$ matrices (see Appendix \ref{Gam}), then we will show how this set of solutions can be transformed into the corresponding set in the \textit{chiral} representation\index{representation!chiral} of the $\gamma$ matrices.  The most useful references for some of the material below are \cite{IZ80}, \cite{T11}, \cite{G00}.\\

\noindent In general, to obtain solutions of the Dirac equation, one would have to solve a set of coupled partial differential equations.  For example, in the standard representation of the $\gamma$ matrices, the Dirac equation \eqref{DE4}, for the $\psi_{i}$ of Eq. \eqref{DE5}  becomes the set of coupled partial differential equations below
\begin{align}
i\partial_{t}\psi_{1}+i\partial_{z}\psi_{3}+i\partial_{x}\psi_{4}+\partial_{y}\psi_{4}-m\psi_{1}&=0,\label{S1}\\
i\partial_{t}\psi_{2}+i\partial_{x}\psi_{3}-\partial_{y}\psi_{3}-i\partial_{z}\psi_{4}-m\psi_{2}&=0,\label{S2}\\
-i\partial_{z}\psi_{1}-i\partial_{x}\psi_{2}-\partial_{y}\psi_{2}-i\partial_{t}\psi_{3}-m\psi_{3}&=0,\label{S3}\\
-i\partial_{x}\psi_{1}+\partial_{y}\psi_{1}+i\partial_{z}\psi_{2}-i\partial_{t}\psi_{4}-m\psi_{4}&=0.\label{S4}
\end{align}

\noindent However in the present case we seek solutions for plane waves of the form
\begin{align}
\psi^{(+)}&=u(p)\,e^{-ip_\mu{}x^\mu{}},\label{PW1.1}\\
\psi^{(-)}&=v(p)\,e^{ip_\mu{}x^\mu{}},\label{PW1.1b}
\end{align}

\noindent where $\psi^{(+)}$ will be the positive energy solutions and $\psi^{(-)}$ the negative energy solutions.  Substituting Eqs. \eqref{PW1.1} and \eqref{PW1.1b} in Eq. \eqref{DE3}, we find the set of algebraic equations below,
\begin{align}
(\slashed{p}-mI)u(p)&=0, \label{PW1.2}\\
(\slashed{p}+mI)v(p)&=0. \label{PW1.2b}
\end{align}

\noindent We note that Eqs. \eqref{PW1.2} and \eqref{PW1.2b} are systems of homogeneous equations for the components of $u(p)$ and $v(p)$.  These systems will have a non-trivial solutions only if
\begin{equation}
\label{PW1.3}
\det(\slashed{p}\pm mI)=0.
\end{equation}

\noindent Equation \eqref{PW1.3} \footnote{Best evaluated with Mathematica for a couple of representations.} gives us the (representation independent) condition
\begin{equation}
\label{PW1.4}
\left(p^{2}-m^{2}\right)^{2}=0,
\end{equation}

\noindent where $p^{2}=\left(p^0{}\right)^{2}-\boldsymbol{p}^{2}\equiv E^{2}-\boldsymbol{p}^{2}$, and therefore Eq. \eqref{PW1.4} may be rewritten as
\begin{equation}
\label{PW1.5}
\left[\left(E-\sqrt{\boldsymbol{p}^{2}+m^{2}}\right)\left(E+\sqrt{\boldsymbol{p}^{2}+m^{2}}\right)\right]^{2}=0.
\end{equation}

\noindent We see that condition \eqref{PW1.3} is satisfied for both $E=\pm\sqrt{\boldsymbol{p}^{2}+m^{2}}$.\\

\noindent Our set of solutions consists of four linearly independent 4-component spinors.\index{spinor!4-component}  We will use $\psi$ for the spinors in the standard representation and $\phi$ for the spinors in the chiral representation (Sec. \ref{C}).  To avoid confusion, we define $\epsilon$ by
\begin{equation}
\label{S4.1}
p_0{}=p^0{}\equiv\epsilon=+\sqrt{(p^x{})^2{}+(p^y{})^2{}+(p^z{})^2{}+m^{2}}
\end{equation}

\noindent For the case of $m\neq 0$ we adopt the normalization of ref. \cite{IZ80} (see Section \ref{norm} below),
\begin{equation}
\label{S5}
N=\sqrt{\frac{\epsilon+m}{2m}}.
\end{equation}

\noindent It is now easy to verify, using Eq. \eqref{PW1.2} with $p_0{}=\epsilon$, that we obtain the two positive energy spinors\index{spinor!positive energy} $u^{(1)}(p)$ and $u^{(2)}(p)$, ($S_z{}=+1/2$ and $S_z{}=-1/2$, respectively) below,\\
\begin{equation}
\label{S6}
u^{(1)}(p)=N\left(\begin{array}{c}1\\\\
0\\\\
 \displaystyle{\frac{p^z{}}{\epsilon+m}}\\\\
 \displaystyle{\frac{p^x{}+ip^y{}}{\epsilon+m}}
\end{array}\right),
\;\;\;\;u^{(2)}(p)=N\left(\begin{array}{c}0\\\\
1\\\\
 \displaystyle{\frac{p^x{}-ip^y{}}{\epsilon+m}}\\\\
 \displaystyle{\frac{-p^z{}}{\epsilon+m}}
\end{array}\right).
\end{equation}\\

\noindent We could repeat the calculations, with $p_0{}=-\,\epsilon$, and obtain two negative energy spinors.\index{spinor!negative energy}   However it is preferable to use the convention adopted in most textbooks (following the Feynman-St\"{u}ckelberg interpretation).   So from Eq. \eqref{PW1.2b}, we obtain the two negative energy spinors\index{spinor!negative energy} $v^{(1)}(p)$ and $v^{(2)}(p)$ below,\\
\begin{equation}
\label{S7}
v^{(1)}(p)=N\left(\begin{array}{c} \displaystyle{\frac{p^z{}}{\epsilon+m}}\\\\
 \displaystyle{\frac{p^x{}+ip^y{}}{\epsilon+m}}\\\\
1\\\\
0
\end{array}\right),
\;\;\;\;v^{(2)}(p)=N\left(\begin{array}{c} \displaystyle{\frac{p^x{}-ip^y{}}{\epsilon+m}}\\\\
 \displaystyle{\frac{-p^z{}}{\epsilon+m}}\\\\
0\\\\
1
\end{array}\right).
\end{equation}\\

\noindent We summarize here some notation and results.  To avoid notational ambiguities we will write when necessary,
\begin{equation}
\label{S7a}
\psi^{(+)(\alpha)}(x)=\left(\begin{array}{c}u^{(\alpha)}_{\,\;1}(p)\\\\
u^{(\alpha)}_{\,\;2}(p)\\\\
u^{(\alpha)}_{\,\;3}(p)\\\\
u^{(\alpha)}_{\,\;4}(p)
\end{array}\right)e^{-ip_\mu{}x^\mu{}},
\end{equation}

\noindent and so on, thus,
\begin{align}
\psi^{(+)(\alpha)}(x)&=u^{(\alpha)}(p)e^{-ip_\mu{}x^\mu{}},\label{S8}\\\nonumber\\
\psi^{(-)(\alpha)}(x)&=v^{(\alpha)}(p)e^{ip_\mu{}x^\mu{}},\;\;\;\;\alpha=(1,2).\label{S9}
\end{align}

\noindent Each member of the set $\psi$ (or $\phi$, Sec. \ref{C}) is an eigenstate of the energy,\index{eigenstate!of energy} and momentum\index{eigenstate!of momentum}.  We adopt the notation below so that Eqs. \eqref{PW1.1} and \eqref{PW1.1b} are:
\begin{equation}
\label{PW1}
\left.\begin{aligned}
&\psi^{(+)(1)},\;\;S_{z}=+\textstyle{\frac{1}{2}}\\
&\psi^{(+)(2)},\;\;S_{z}=-\textstyle{\frac{1}{2}}
\end{aligned}\;\;\right\}
\qquad \text{positive energy}
\end{equation}
\begin{equation}
\label{PW2}\
\left.\begin{aligned}
&\psi^{(-)(1)},\;\;S_{z}=+\textstyle{\frac{1}{2}}\\
&\psi^{(-)(2)},\;\;S_{z}=-\textstyle{\frac{1}{2}}
\end{aligned}\;\;\right\}
\qquad \text{negative energy}
\end{equation}\\

\noindent We note that since $i\partial_t{}\psi=H\psi=E\psi$, we have
\begin{align}
i\partial_{t}\psi^{(+)(\alpha)}(x)&=p_0{}\,\psi^{(+)(\alpha)}(x),\label{S10a}\\
i\partial_{t}\psi^{(-)(\alpha)}(x)&=-p_0{}\,\psi^{(-)(\alpha)}(x).\label{S11a}
\end{align}

\noindent Likewise,
\begin{align}
-i\partial_{z}\psi^{(+)(\alpha)}(x)&=p^z{}\,\psi^{(+)(\alpha)}(x),\label{S10b}\\
-i\partial_{z}\psi^{(-)(\alpha)}(x)&=-p^z{}\,\psi^{(-)(\alpha)}(x).\label{S11b}
\end{align}

\begin{remark}\label{sign}\index{Remarks} Note that although $p_0{}>0$ in both Eqs. \eqref{S8} and \eqref{S9}, some of the \textup{3}-momenta have opposite directions.  
\end{remark}

\subsection[Chiral representation set]{Chiral representation set}\label{CR}

\noindent There is a fundamental theorem by Pauli\index{Pauli theorem} which states that for any two four-dimensional representations of the Dirac $\gamma$ matrices there exists a nonsingular $4\times 4$ matrix $U$, such that $\gamma^{\prime\,A}=U\gamma^A{}U^{-1}$.  Moreover if $\gamma^{0\dagger}=\gamma^0{},\;\gamma^{K\dagger}=-\gamma^K{}$, for $K=1,2,3$, the matrix $U$ is unitary (e.g., see \cite{G00}).  The unitary matrix $U$\index{unitary matrix} below relates the standard representation\index{representation!standard} to our version of the chiral representation,\index{representation!chiral} Eqs. \eqref{D6}.
\begin{equation}
\label{C1}
 U=\frac{1}{\sqrt{2}}\left(
 \begin{matrix}
     I_{2} &  I_{2}\\
     \rule{0in}{2ex}
    -I_{2} & I_{2}\\
 \end{matrix}
 \right).
\end{equation}

\noindent We have that
\begin{equation}
\label{C2}
\gamma^{\mu}_{\;(chiral)}=U\gamma^{\mu}_{\;(stand)}U^{-1},
\end{equation}

\noindent compare Eqs. \eqref{D4} and \eqref{D6}.  Using Eq. \eqref{DE1} we write
\begin{equation}
\label{C3}
iU\gamma^{\mu}U^{-1}U\partial_{\mu}\psi-mU\psi=0\,,
\end{equation}

\noindent so instead of Eqs. \eqref{S1} - \eqref{S4}, we now have,
\begin{align}
-i\partial_{t}\phi_{3}+i\partial_{z}\phi_{3}+i\partial_{x}\phi_{4}+\partial_{y}\phi_{4}-m\phi_{1}&=0,\label{C4}\\
i\partial_{x}\phi_{3}-\partial_{y}\phi_{3}-i\partial_{t}\phi_{4}-i\partial_{z}\phi_{4}-m\phi_{2}&=0,\label{C5}\\
-i\partial_{t}\phi_{1}-i\partial_{z}\phi_{1}-i\partial_{x}\phi_{2}-\partial_{y}\phi_{2}-m\phi_{3}&=0,\label{C6}\\
-i\partial_{x}\phi_{1}+\partial_{y}\phi_{1}-i\partial_{t}\phi_{2}+i\partial_{z}\phi_{2}-m\phi_{4}&=0,\label{C7}
\end{align}

\noindent where
\begin{equation}
\label{C8}
\phi=U\psi.
\end{equation}

\noindent We use Eq. \eqref{C8} and Eqs. \eqref{S8}, \eqref{S9}, and obtain the chiral positive\index{spinor!positive energy} and negative energy\index{spinor!negative energy} solutions below,\\
\begin{align}
&\phi^{(+)(1)}=\frac{N}{\sqrt{2}}\left(\begin{array}{c}\displaystyle{1+\frac{p^z{}}{\epsilon+m}}\\\\
\displaystyle{\frac{p^x{}+i\,p^y{}}{\epsilon+m}}\\\\
\! \! \! \displaystyle{-1+\frac{p^z{}}{\epsilon+m}}\\\\
\displaystyle{\frac{p^x{}+i\,p^y{}}{\epsilon+m}}
\end{array}\right)
e^{-ip_\mu{}x^\mu{}},\;\;\;
\phi^{(+)(2)}=\frac{N}{\sqrt{2}}\left(\begin{array}{c}\displaystyle{\frac{p^x{}-i\,p^y{}}{\epsilon+m}}\\\\
\displaystyle{1-\frac{p^z{}}{\epsilon+m}}\\\\
\! \! \! \;\displaystyle{\frac{p^x{}-i\,p^y{}}{\epsilon+m}}\\\\
\displaystyle{-1-\frac{p^z{}}{\epsilon+m}}
\end{array}\right)
e^{-ip_\mu{}x^\mu{}},\nonumber\\\nonumber\\\label{C9}\\
&\phi^{(-)(1)}=\frac{N}{\sqrt{2}}\left(\begin{array}{c}\displaystyle{1+\frac{p^z{}}{\epsilon+m}}\\\\
\displaystyle{\frac{p^x{}+i\,p^y{}}{\epsilon+m}}\\\\
\! \! \! \displaystyle{1-\frac{p^z{}}{\epsilon+m}}\\\\
\displaystyle{-\frac{p^x{}+i\,p^y{}}{\epsilon+m}}
\end{array}\right)
e^{ip_\mu{}x^\mu{}},\;\;\;\;\;\;\;\;
\phi^{(-)(2)}=\frac{N}{\sqrt{2}}\left(\begin{array}{c}\displaystyle{\frac{p^x{}-i\,p^y{}}{\epsilon+m}}\\\\
\displaystyle{1-\frac{p^z{}}{\epsilon+m}}\\\\
\! \! \!\displaystyle{\frac{-p^x{}+i\,p^y{}}{\epsilon+m}}\\\\
\displaystyle{1+\frac{p^z{}}{\epsilon+m}}
\end{array}\right)e^{ip_\mu{}x^\mu{}}.\nonumber
\end{align}\\

\begin{remark}\label{chiral}\index{chiral}  It is worth pointing out that if we use the notation of Eq. \eqref{S7a} for the above \textup{(}chiral\textup{)} wavefunctions, and substitute these functions in the Dirac equations Eqs. \eqref{C4} - \eqref{C7}, and let $p^x{}=p^y{}=p^z{}=0$, we will find that for the $\phi^{(+)(i)}$, $i=1,2$, the equations are satisfied with $p^0{}=\epsilon=m$, while for the $\phi^{(-)(i)}$, the equations are satisfied with $p^0{}=-m$.  
\end{remark}

\noindent The chiral operator\index{chiral operator} $\gamma^{5}$, Eq. \eqref{D8}, in the chiral representation\index{representation!chiral} is
\begin{equation}
\label{C11}
\gamma^{5}=\left(
 \begin{matrix}
     I_{2} &  0\\
     \rule{0in}{2ex}
    0& -I_{2}\\
 \end{matrix}
 \right).
\end{equation}

\noindent We introduce the \textit{bispinor}\index{bispinor} $\chi$, \cite{S13} below to represent the spinors in Eqs. \eqref{C9},
\begin{equation}
\label{C12}
\chi=\left(\begin{array}{c}\chi_{R}\\
\chi_{L}
\end{array}\right),
\end{equation}

\noindent where each entry is a 2-component spinor\index{spinor!2-component}.  We then have
\begin{equation}
\label{C13}
\gamma^{5}\chi=\left(\begin{array}{c}+\chi_{R}\\
-\chi_{L}
\end{array}\right),
\end{equation}

\noindent $\chi_{R}$ being right-handed\index{spinor!right-handed} and $\chi_{L}$ left-handed.\index{spinor!left-handed}

\begin{remark}\label{eigen}\index{eigen}  As mentioned above, the results given in Eqs. \eqref{S10a} - \eqref{S11a} for the $\psi$'s also hold for the $\phi$'s of Eq. \eqref{C9}.  
\end{remark}

\subsection[Normalization of $\psi$]{Normalization of $\psi$}\label{norm}

\noindent We follow Itzykson and Zuber \cite{IZ80} and adopt the Lorentz invariant normalizations\index{normalization!Lorentz invariant}
\begin{align}
\bar{u}^{(\alpha)}(p)u^{(\beta)}(p)&=\delta^{\,\alpha\beta}\,,&\bar{u}^{(\alpha)}(p)v^{(\beta)}(p)=0\,,\label{N8}\\
\bar{v}^{(\alpha)}(p)v^{(\beta)}(p)&=-\delta^{\,\alpha\beta}\,,&\bar{v}^{(\alpha)}(p)u^{(\beta)}(p)=0\,.\label{N9}
\end{align}

\begin{remark}\label{normal}\index{Remarks}  We remark that different authors adopt different normalizations for the Lorentz invariant product $\bar{\psi}\psi$ \textup{(}the invariance itself is a little tedious to show, see, e.g., the last page of \textup{\cite{T11}}\textup{)}.
\end{remark}

\noindent The normalization factor\index{normalization!factor} $N$ for the plane waves, given by Eq. \eqref{S5}, follows from Eqs. \eqref{N8}, \eqref{N9}, using the solutions \eqref{S6} and \eqref{S7}.\\

\noindent For easy reference for the proofs to follow, we write again some of the formulas derived above:
\begin{align}
\left(\slashed{p}-mI\right)u^{(\alpha)}(p)=0,\;\;\;\;\bar{u}^{(\alpha)}(p)\left(\slashed{p}-mI\right)=0,\label{N9.1}\\
\left(\slashed{p}+mI\right)v^{(\alpha)}(p)=0,\;\;\;\;\bar{v}^{(\alpha)}(p)\left(\slashed{p}+mI\right)=0.\label{N9.2}
\end{align}

\noindent We shall derive an expression for $j^\mu{}=\bar{\psi}\gamma^{\mu}\psi$.  In the derivation we make use of Eqs. \eqref{N8} - \eqref{N9.2}.  For the positive energy solutions we have,
\begin{align}
\bar{\psi}^{(+)(\alpha)}\gamma^{\mu}\psi^{(+)(\beta)}&=\bar{u}^{(\alpha)}(p)\gamma^\mu{}u^{(\beta)}(p),\nonumber\\
&=\frac{1}{2}\left[\left(\bar{u}^{(\alpha)}\gamma^\mu{}\right)u^{(\beta)}+\bar{u}^{(\alpha)}\left(\gamma^\mu{}u^{(\beta)}\right)\right],\nonumber\\
&=\frac{1}{2m}\left[\left(\bar{u}^{(\alpha)}m\gamma^\mu{}\right)u^{(\beta)}+\bar{u}^{(\alpha)}\left(\gamma^\mu{}mu^{(\beta)}\right)\right],\nonumber\\
&=\frac{1}{2m}\left[\left(\bar{u}^{(\alpha)}\slashed{p}\gamma^\mu{}\right)u^{(\beta)}+\bar{u}^{(\alpha)}\left(\gamma^\mu{}\slashed{p}u^{(\beta)}\right)\right],\nonumber\\
&=\frac{1}{2m}\left[\bar{u}^{(\alpha)}\{\slashed{p},\gamma^\mu{}\}u^{(\beta)}\right],\nonumber\\
&=\frac{1}{2m}\left[\bar{u}^{(\alpha)}p_\nu{}\{\gamma^\nu{},\gamma^\mu{}\}u^{(\beta)}\right],\nonumber\\
&=\frac{1}{2m}\left[\bar{u}^{(\alpha)}p_\nu{}2\eta^{\nu\mu}Iu^{(\beta)}\right]=\frac{p^\mu{}}{m}\delta^{\alpha\beta}\label{N10}\,.
\end{align}

\noindent Repeating this derivation for the negative energy solutions we get
\begin{align}
\bar{\psi}^{(-)(\alpha)}\gamma^{\mu}\psi^{(-)(\beta)}&=\bar{v}^{(\alpha)}(p)\gamma^\mu{}v^{(\beta)}(p),\nonumber\\
&=-\frac{1}{2m}\left[\bar{v}^{(\alpha)}\{\slashed{p},\gamma^\mu{}\}v^{(\beta)}\right],\nonumber\\
&=-\frac{1}{2m}\left[\bar{v}^{(\alpha)}p_\nu{}2\eta^{\nu\mu}Iv^{(\beta)}\right]=\frac{p^\mu{}}{m}\delta^{\alpha\beta}\label{N11}\,.
\end{align}

\noindent It is important to show that positive and negative energy states are mutually orthogonal if we consider states with opposite energies \textit{but the same \textup{3}-momentum}.  We recall Eqs. \eqref{S10a}, \eqref{S11a} and write explicitly
\begin{align}
\psi^{(+)(\alpha)}(x)&=u^{(\alpha)}(p)e^{-i\left(p^0{}x^0{}-p^i{}x^i{}\right)},\label{N12}\\
\psi^{(-)(\beta)}(x)&=v^{(\beta)}(q)e^{i\left(p^0{}x^0{}+p^i{}x^i{}\right)}\,,\label{N13}
\end{align}

\noindent where the vector (not the covector $p_\mu{}$) momenta are $p=(p^0{},\boldsymbol{p})$, $q=(p^0{},-\boldsymbol{p})$, respectively, see Remark \ref{sign}.  Therefore, using again Eqs. \eqref{N8} - \eqref{N9.2}, we have
\begin{align}
\bar{\psi}^{(-)(\beta)}\psi^{(+)(\alpha)}&=e^{-2ip^0{}x^0{}}\bar{v}^{(\beta)}(q)\gamma^{0}u^{(\alpha)}(p),\nonumber\\
&=\frac{1}{2m}e^{-2ip^0{}x^0{}}\left[\left(\bar{v}^{(\beta)}(q)m\right)\gamma^{0}u^{(\alpha)}(p)+\bar{v}^{(\beta)}(q)\gamma^{0}\left(mu^{(\alpha)}(p)\right)\right],\nonumber\\
&=\frac{1}{2m}e^{-2ip^0{}x^0{}}\bar{v}^{(\beta)}(q)\left(-\slashed{q}\gamma^{0}+\gamma^{0}\slashed{p}\right)u^{(\alpha)}(p)=0.\label{N14}
\end{align}

\noindent Showing the last step above requires care!\\

\noindent We now define the scalar product with the standard delta function normalization\index{normalization!delta function} for free particles,
\begin{equation}
\label{N15}
\left(\psi^{(\alpha)}_{\boldsymbol{p}^{\prime}}\vert \psi^{(\beta)}_{\boldsymbol{p}}\right)=\int\bar{\psi}^{(\alpha)}_{\boldsymbol{p}^{\prime}}\gamma^0{}\,\psi^{(\beta)}_{\boldsymbol{p}}d^{3}x=\delta^{\alpha\beta}\delta^{3}\left(\boldsymbol{p}^{\prime}-\boldsymbol{p}\right)\,,
\end{equation}

\noindent where the delta function is given by
\begin{equation}
\label{N16}
\delta^{3}\left(\boldsymbol{p}^{\prime}-\boldsymbol{p}\right)=\frac{1}{(2\pi)^{3}}\int e^{i\left(\boldsymbol{p}^{\prime}-\boldsymbol{p}\right)\boldsymbol{r}}d^{3}x,\;\;\;\;\boldsymbol{r}=\left(x^1{},x^2{},x^3{}\right).
\end{equation}

\noindent As a consequence of result \eqref{N14} we see that if, in the integrand of \eqref{N15}, the $\psi$'s have opposite energies, the result of the integration is zero (either from Eq. \eqref{N14} or from the $\delta^{3}\left(\boldsymbol{p}^{\prime}-\boldsymbol{p}\right)$).

\begin{remark}\label{minor}\index{Remarks}  Since $\epsilon=+\sqrt{\boldsymbol{p}^{2}+m^{2}}$, if the absolute value of the $\boldsymbol{p}$'s is the same, then the absolute value $\epsilon$ of the energy is the same.
\end{remark}

\noindent At this point we must introduce another normalization factor,\index{normalization!factor} $\tilde{N}$, required by the integration over space (we could have introduced $\tilde{N}$ in Eqs. \eqref{S8} and \eqref{S9} but it would only have complicated the writing).  Effectively this amounts to re-defining the $\psi$'s as follows, 
\begin{align}
\psi^{(+)(\alpha)}(x)&=\tilde{N}u^{(\alpha)}(p)e^{-ip_\mu{}x^\mu{}},\label{N17}\\
\psi^{(-)(\alpha)}(x)&=\tilde{N}v^{(\alpha)}(p)e^{ip_\mu{}x^\mu{}}.\label{N18}
\end{align}

\noindent The calculation of $\tilde{N}$ is simpler if we consider ``box normalization''\index{normalization!box} with periodic boundary conditions\index{boundary conditions!periodic} in a volume $V$.  Then, instead of Eq. \eqref{N15} we have
\begin{equation}
\label{N19}
\left(\psi^{(\alpha)}_{\boldsymbol{p}^{\prime}}\vert\psi^{(\beta)}_{\boldsymbol{p}}\right)=\int\bar{\psi}^{(\alpha)}_{\boldsymbol{p}^{\prime}}\gamma^0{}\,\psi^{(\beta)}_{\boldsymbol{p}}d^{3}x=\delta^{\alpha\beta}\delta_{\boldsymbol{p}^{\prime}\boldsymbol{p}}\,,
\end{equation}

\noindent thus
\begin{equation}
\label{N20}
\int\bar{\psi}^{(\alpha)}_{\boldsymbol{p}}\gamma^0{}\,\psi^{(\beta)}_{\boldsymbol{p}}d^{3}x=\int j^0{}d^{3}x=\frac{p^0{}}{m}\delta^{\alpha\beta}\tilde{N}^{2}\,V,
\end{equation}

\noindent and so
\begin{align}
\psi^{(+)(\alpha)}(x)&=\frac{1}{\sqrt{V}}\sqrt{\frac{m}{p^0{}}}u^{(\alpha)}(p)e^{-ip_\mu{}x^\mu{}},\label{N21}\\
\psi^{(-)(\alpha)}(x)&=\frac{1}{\sqrt{V}}\sqrt{\frac{m}{p^0{}}}v^{(\alpha)}(p)e^{ip_\mu{}x^\mu{}},\label{N22}
\end{align}

\noindent where we recall that $p^0{}=\epsilon$.  If we had used Eq. \eqref{N15} instead of Eq. \eqref{N19}, we would have to do the replacement (cf. ref. \cite{S67}),
\begin{equation}
\label{N23}
\frac{1}{\sqrt{V}}\rightarrow\frac{1}{(2\pi)^{3/2}}.
\end{equation}

\end{appendices}

\printindex

\end{document}